\documentclass[aps,prl,showpacs,floatfix]{revtex4}
\usepackage{graphicx}
\usepackage[fleqn]{amsmath}
\usepackage{bm}
\usepackage{epsf,epsfig,graphics}
\usepackage{makeidx}
\usepackage{latexsym}
\usepackage{amssymb}
\usepackage{mathrsfs}
\usepackage{amsbsy}
\usepackage{subfigure}
\newcommand*\colvec[3][]{\begin{pmatrix}\ifx\relax#1\relax\else#1\\\fi#2\\#3\end{pmatrix}}

\newcommand{\figws}{8cm}
\newcommand{\figwss}{5.cm}

\usepackage{color}
\usepackage{morefloats}
\usepackage{CJKutf8}
\DeclareMathAlphabet\mathbfcal{OMS}{cmsy}{b}{n}

\begin{document}
\title{ Regulatory Feedback Effects on Tissue Growth Dynamics in a Two-Stage Cell Lineage Model}
\author{Mao-Xiang Wang\footnote{Email:
wangmx@njust.edu.cn}$^{(1,2,3)}$\begin{CJK*}{UTF8}{bsmi}(王茂香)\end{CJK*}, Arthur Lander$^{(2,3)}$,  and Pik-Yin Lai\footnote{Email:
 pylai@phy.ncu.edu.tw}$^{(4)}$\begin{CJK*}{UTF8}{bsmi}(黎璧賢)\end{CJK*}}
\affiliation{$^{(1)}$School of Science, Nanjing University of Science and Technology,  Nanjing 210094, China\\
$^{(2)}$ Department of Developmental and Cell Biology, University of California, Irvine, Irvine, California, USA  \\
$^{(3)}$   Center for Complex Biological Systems, University of California, Irvine, Irvine, California, USA \\
  $^{(4)}$  Dept. of Physics and Center for Complex Systems, National Central
University, Chungli District, Taoyuan City, Taiwan 320, R.O.C.}
\date[] {\bf  \today}
\begin{abstract}
Identifying the mechanism of intercellular feedback regulation is critical for the basic understanding of tissue growth control  in organisms. In this paper,  we analyze a  tissue growth model consisting of a single lineage of two cell types regulated by  negative feedback signalling molecules that undergo spatial diffusion. By deriving the fixed points for the uniform steady states and carrying out linear stability analysis,   phase diagrams are obtained analytically for arbitrary parameters of the model.   Two different generic  growth modes are found:  blow-up growth and final-state controlled growth which  are governed by the  non-trivial  fixed point and the trivial fixed point respectively, and  can be sensitively switched by varying the negative feedback regulation on the proliferation of the stem cells. Analytic expressions for the characteristic   time scales  for these two growth modes are also derived. Remarkably, the trivial and non-trivial uniform steady states can coexist and a sharp transition occurs in the bistable regime as the relevant parameters are varied.  Furthermore, the bi-stable growth properties allows for the external control to switch between these two growth modes. In addition, the condition for an early accelerated growth followed by a retarded growth can be derived. These analytical results are further  verified by  numerical simulations and provide insights on the growth behavior of the tissue.  Our results are also discussed in the light of possible  realistic biological experiments and tissue growth control strategy. Furthermore, by external feedback control of the concentration of regulatory molecules, it is possible to achieve a desired growth mode, as demonstrated with an analysis of boosted growth, catch-up growth and the design for the target of a linear growth dynamic.  
\end{abstract}
 \maketitle

\section{Introduction}
  Biological functions are carried out by organs composed of tissues of specific architecture and sizes in high-level multi-cellular complex organisms. In the developmental stage, the  growth of tissue is governed by the interplay of cell proliferation, differentiation, and  cell apoptosis\cite{wikicell} and regulated by feedback signals for  proliferation and/or differentiation down the lineage\cite{tracking} so as to ensure a normal pathway leading to an appropriate tissue size\cite{Lander2009}. 
Such feedback regulations are often achieved by cell-cell communications, such as via quorum sensing\cite{QS, WY2011, Laiward2017} in which bacteria are able to sense  the cell density and regulate their proliferation processes accordingly. 
The ability to detect signalling chemicals is also essential for cell differentiation in development\cite{yamanaka2014}, for example concentration gradients of BMP and Wnt along two orthogonal axes are responsible for both dorsal-ventral and anterior-posterior axes formation\cite{Wnt}. Recently, signalling  molecules that control the output of multistage cell lineages have been explored  in the olfactory epithelium of mice\cite{Lander2010}, revealing that the spatial distribution of diffusive signaling molecules (including GDF11, Activin $\beta$B and Follistatin)  regulate the proliferation of each cell type within the lineage and help to generate tissue stratification through controlling the spatial distribution of these signaling molecules. Inhibitory feedback regulation from signaling chemical acting on the proliferating cells in general will suppress the cell population and hence achieve in the control of the tissue sizes.           
       
Cell lineage is the basic unit of tissue and organ formation. The molecular mechanisms underlying the control of growth and regeneration of tissues and organs are subjects of fundamental biological interests as well as medical concerns\cite{tracking}.  Recent experiments have shown that in the mouse Olfactory Epithelium, a secreted molecule, GDF11 produced by terminally differentiated (TD) cells, feeds back upon intermediate progenitors,  and  together  with another molecule that feeds back upon stem cells (Activin $\beta$B produced by TD cells),  creating a dual feedback loop\cite{Lander2009}. Based on the above observations, a logic proliferation control model has been built (the ODE model in Fig. 1) and the theoretical results indicating that negative feedback on self-renewal indeed stabilizes the exponentially growth of the cell system,  thus producing a steady state tissue size. Such feedback regulations are often carried out through diffusive molecules\cite{Laiward2017}, such as morphogens, growth factors, cytokins and chemokines. A spatial model of multistage cell lineage and negative feedback regulation indicates that tissue stratification can be generated and maintained through controlling  the spatial distribution of diffusive signaling molecules\cite{Lander2010,yeh2018}. A mathematical model consisting of a short-range activation of Wnt and a long-range inhibition with modulation of BMP signals in a growing tissue of cell lineage can account for the formation, regeneration and stability of intestinal crypts\cite{Lander2012}. In addition, a particular feedback architecture in which both positive and negative diffusible signals act on stem cells can lead to the appearance of bi-modal growth behaviors, and resulting in some kind of self-organizing morphogenesis\cite{Lander2016}. Up to now, numerous biological experiments and modeling have been extensively studied from a mechanistic perspective, but little efforts have been directed toward the understanding of the logic of control. In this paper, we attempt to address the above questions through  theoretical analysis on a built model and investigate the fundamental control principles that shape the general architectures of these biological systems.     
       
By incorporating spatial diffusive regulatory molecules, we examine  the effects of feedback regulation via such signaling molecules on the cell growth and tissue size of a simple cell lineage model. Interesting,  in some parameter regime, a bistable regime of  two uniform steady states can coexist.
In contrast to previous studies on feedback-driven morphogenesis which require both positive and negative feedbacks to achieve the bi-stability, in our model mere negative feedback regulation on the proliferation of stem cells can realize the bi-modal growth. While most previous studies relies on numerical solution of the model equations, here we manage to carrying out  a thorough  theoretical calculations that lead to analytic results on the tissue growth dynamics and growth stability. Our theoretical results holds for rather general negative feedback function and bistability occurs if the feedback suppression is sufficiently sensitive.
In particular, the feedback is modeled by a Hill function and the full phase diagram can be obtained analytically together with the phase boundaries for arbitrary values of the Hill coefficient, regulation strengths and other parameters of the model.   The analytical results are further verified by direct numerical solution of the model equations. Possible applications in tissue growth engineering strategies and control such as switching of growth modes by external pulse control, engineered linear growth, catch-up growth and the timing precision in growth boosting, are also proposed and discussed.

\section{Cell Lineage Model with Negative feedback control}
 Cell lineage denotes the developmental history of a tissue or organ from the fertilized embryo. An un-branched uni-directional cell lineages may be produced by a sequence of differentiation that begin with a stem cell (SC), progress through some number of self-renewing progenitor stages, and end with one or more terminal differentiated (TD) cells\cite{Lander2009}. 
 On the other hand, homeostatic control is an important goal for regulation and feedback mechanism to maintain the stability in biological tissues\cite{Komarova}. Furthermore, negative feedback regulations occur much more often  than positive feedbacks so as to maintain a well-controlled growth and development in biological systems. For example, negative feedback regulates tissue sizes and enhances the regeneration. In the model of mammalian olfactory epithelium, the tissue contains  SCs, transit amplifying (TA) cells and TD cells. Each cell can potentially secret regulatory factors, and respond to factors secreted by other cell types.  With suitable negatively regulated processes, the regulatory molecules can avoid the fate of uncontrolled growth and also can achieve the target cell population and tissue size that is biologically appropriate. 

In some cell lineages, the TD cells constantly turn over, as occurs in hematopoietic, epidermal and many epithelial lineages. The balance  between the turnover  and production of the TD cells is essential to sustain homeostasis, which can be viewed as achieving a steady state\cite{Lander2010} in the cell dynamics. However, not all tissues can reach a dynamically steady state, in such a case the  TD cells last for the entire lifetime of the organism (e.g. in the nervous system), with the SC either disappearing or becoming quiescent. Such a scenario is referred   as the ``final-state" of the system\cite{Lander2016}. 

\begin{figure}[h]
  \subfigure[]{\includegraphics*[width=\figwss]{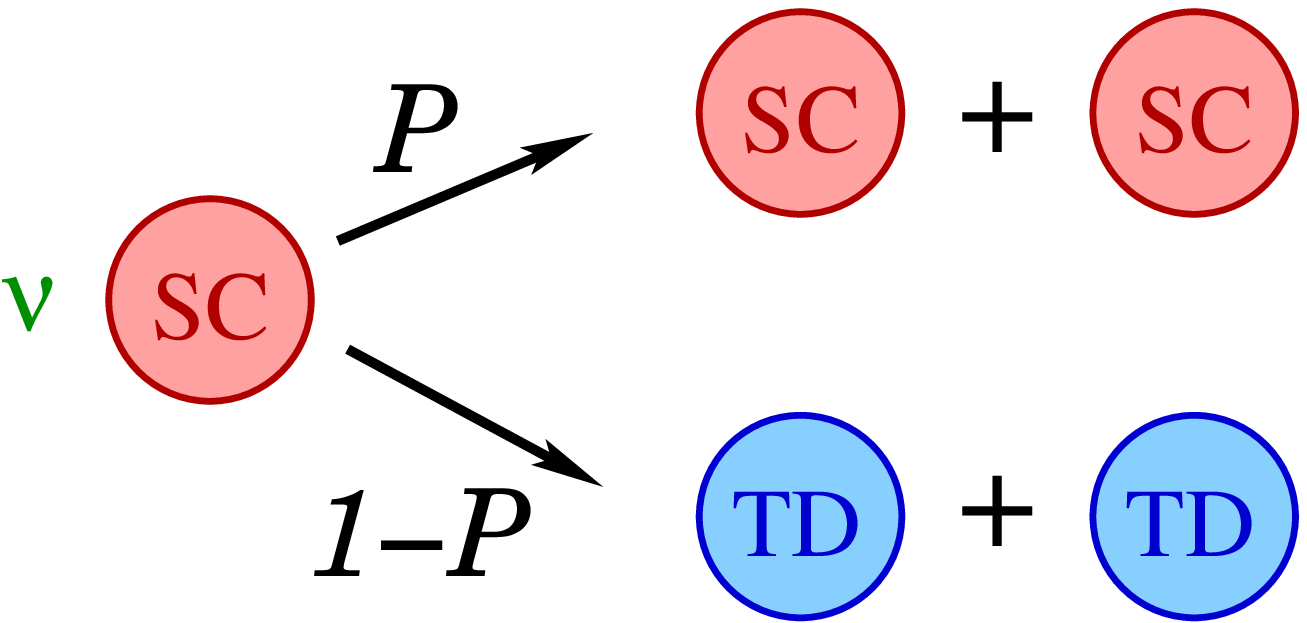}}
    \subfigure[]{\includegraphics*[width=\figws]{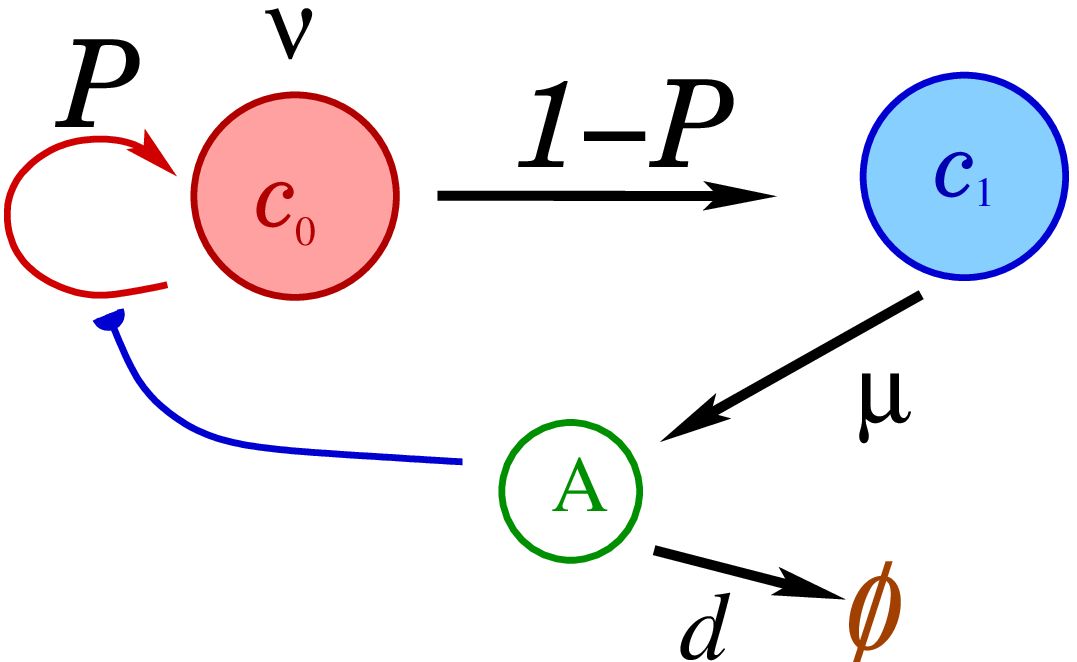}}
  \caption{ Schematic picture showing a two-stage cell lineage model with negative feedback. $c_0$ and $c_1$ are the concentration of stem cell (SC) and terminally differentiated (TD) cell respectively. (a) SC replicates with a rate of $\nu$ and has a probability ${\cal P}$  are  of duplicating itself and $1-{\cal P}$ of differentiating into two TD cells.  (b) The  molecule $A$, which negative regulates ${\cal P}$,  is produced by the TD cells at a rate $\mu$, and decays with a rate $d$.
   }\label{schematic}
\end{figure}

One of the simplest  unbranched two-cell lineage systems  is depicted schematically in Fig.\ref{schematic}. The existence and uniqueness as well as local and global stability of steady states in the corresponding ODE system of multistage cell lineages generalization have been established\cite{Lander2008}. Tissue stratification and regeneration of intestinal crypts can be successfully modeled by considering spatial advection of cells and regulating molecules, and it is suggested that the turnover of TD cell is necessary to keep a stable dynamic  equilibrium in the system. 
But for the case of final-state tissues, the system is not maintained at some dynamical equilibrium due to the lack of TD cell death (as shown in Fig. \ref{schematic}b), but the population of the TD cell stops due to the extinction of the SC or the SC  becomes inactive. Bone, cartilage, retina and most of the brain are such final-state tissues. Other organs, such as liver, the turnover is so slow that it can be taken to be effectively final-state tissues from the viewpoint of development. Essentially all cases of tight developmental size control over long distance involve final-states. So this mode of growth control is fundamentally different from the checkpoint control, in which the control occurs only near its target state. For the growing tissue that is determined by its final-state, it is best to take the control early (when the dimension of tissue is much less than the decay length of diffusible molecules) and employ a high feedback gain to realize an effective control.  For simplicity and the purpose of illustrating the ideas, we  consider a simple model of two-cell lineage to investigate the strategy of controlling a final-state system. 

 It has been shown in \cite{mxwang,mxReg} that in the continuum limit, different cell  types in different stages of a lineage together with the diffusion of regulatory molecules 
 can be modeled by coupled PDEs of the cell densities and regulatory molecule concentrations. And for the simple case of a two-stage lineage model with cell and molecule advection, the cell density and feedback molecule concentration can be modeled   as\cite{Lander2010}
    \begin{eqnarray}
      \frac{\partial c_0}{\partial t}+ \frac{\partial (V c_0)}{\partial z} &=&\nu c_0(2{\cal P}(A)-1)\nonumber\\
      \frac{\partial c_1}{\partial t}+ \frac{\partial (V c_1)}{\partial z} &=&2\nu c_0(1-{\cal P}(A))\label{PDEG1}\\
      \frac{\partial A}{\partial t}+ \frac{\partial (VA)}{\partial z} &=&D \frac{\partial^2 A}{\partial z^2}+ \mu c_1-A d\nonumber
       \end{eqnarray}
 where  ${\cal P}(A) $  is the stem cell proliferation probability which is a decreasing function (to model negative feedback) of the concentration of regulating molecules $A$. The stem cell differentiates with probability $1- {\cal P}(A)$.
In this two-stage cell lineage in one-dimension, $c_0$ denotes the concentration of stem cell and $c_1$ is the concentration of TD cell.  $\nu$ is the cell cycle rate multiplied by $\ln 2$. $V$ represents the tissue growth velocity driven by the proliferation and differentiation of the cells. $\mu$ and  $d$ represent the secretion  and decay rates of molecules respectively. $D$ is the effective diffusion constant of the molecules. Note that the asymmetric cell division  SC$\to$SC+TD, is not included in the model, which can affect the total stem cell balance mechanism if such a pathway is significant.  But actually even if the above asymmetric division is included, the model can still be reduced\cite{SCTD} to an effective form represented as in Fig. \ref{schematic}a by redefining ${\cal P}$.

It is worth to note that  the TD cells in our model will accumulate rather than being shed and the final-state will be a tissue consisting of only the TD cell resulting in practice a ``once in a life-time” growth. For  a strict final-state system, the TD cells can never turn over. However, if  the turn over rate of TD cells is slow, the final-state  will better describe  the lineage dynamics on time scales that are short relative to the TD cell lifespan. This can serve to describe the fast growth developmental stage in which the growth rate of TD cells is much greater than their  death rate. Since tissue morphogenesis often occurs much more rapidly than the TD cell lifespans, final-state models may thus be inherently better and convenient for lineage dynamics during morphogenesis, even in self-renewing tissues.

For further explicit theoretical calculations, we shall adopt the following Hill function form for the stem cell proliferation probability
\begin{equation}
{\cal P}(A)=\frac{p}{1+(\gamma A)^m},\label{PA}
\end{equation}
where $\gamma$ is the regulation strength and $m$ is Hill coefficient which is a positive number and usually taken to be an integer, $p$ represents the maximal replication probability. The Hill function form in (\ref{PA}) is employed to describes a sharp decrease (if $ m >1$) in ${\cal P}$ as a function of the concentration of the regulatory agent to model a rapid switching off of proliferation when A exceeds some characteristic value. It is rather common to model the feedback regulation by cooperative binding of several regulatory proteins on some binding sites, which can be treated by statistical mechanical means and will lead to a Hill function form\cite{PBoC}.

Since real tissue must grow outward into physical space, it will displace (advect) both the cells and diffusible molecules to potentially different extent at different locations. Moreover, molecules that mediate regulatory feedback will naturally form spatial gradients, and feedback molecules can be considered to diffuse freely among the cells. In the study of the regeneration of intestinal crypt in which the combination of positive and negative feedback is considered, it was suggested a reaction-diffusion mechanism with a short-range activation plus a long-range inhibition can lead to  Turing pattern formation\cite{Lander2012}. In the study of feedback-driven morphogenesis with positive and negative feedback signals,  a bi-modal growth behavior was reported\cite{Lander2016}. Positive and negative feedback certainly exist in realistic biological systems, but negative feedback may play a more dominant role for maintaining homoeostasis. Here we focus on  the mechanism  in which the stem cell proliferation is only regulated by negative feedback  of different strengths. The effects of the secretion and death rates of the  feedback molecules  in realizing different growth behavior and hence in the  control of  tissue sizes are also incorporated in our model. As  will be described below, the growth mode can switch merely by  changing the negative feedback in the absence of any positive regulation.

\section{Analytical Results: Bistability and phase diagrams}
 By choosing the  time and space in   units of $1/d$ and $\sqrt{D/d}$(the characteristic decay length) respectively, the number of parameters in the governing equations  in (\ref{PDEG1}) and (\ref{PA}) can reduced to only four: ${\tilde \mu}\equiv\frac{\gamma\mu}{d}$, ${\tilde \nu}\equiv\frac{\nu}{d}$, $m$ and $p$. The numerical results associated with times and lengths presented in Sec. V are all with the above natural units.
 
Assuming the two types of cells fill up the whole space in which the tissue is occupying, one has the constraint $c_0 + c_1 = 1 $. Thus Eqs. (\ref{PDEG1})  can be simplified to:
   \begin{eqnarray}
      \frac{\partial c_0}{\partial t}+ \frac{\partial (V c_0)}{\partial z} &=&\nu c_0(2{\cal P}(A)-1)\nonumber\\
           \frac{\partial A}{\partial t}+ \frac{\partial (VA)}{\partial z} &=&D \frac{\partial^2 A}{\partial z^2}+ \mu (1-c_0)-A d\label{PDE2}\\
         \frac{\partial V}{\partial z}&=&\nu c_0.\nonumber 
         \end{eqnarray}
(\ref{PDE2}) can be rewritten to give
\begin{eqnarray}
 \frac{\partial c_0}{\partial t}&=& \nu c_0[2{\cal P}(A)-1-c_0]-V(z)\frac{\partial c_0}{\partial z}\label{PDE3}\\
 \frac{\partial A}{\partial t}&=& \mu(1-c_0)-(\nu c_0+d)A-V(z)\frac{\partial A}{\partial z}+D \frac{\partial^2 A}{\partial z^2}\label{PDE4}\\
 V(z)&=&\nu\int_0^z c_0(x,t) dx.\label{Vz2}
\end{eqnarray}         
         The spatially homogeneous solution, or the uniform state dynamics of $c_0(t)$ and $A(t)$, is given by
         putting the spatial gradients in (\ref{PDE3}) and (\ref{PDE4}) to zero, 
   \begin{eqnarray}
 \frac{d c_0}{dt}&=& \nu c_0[2{\cal P}(A)-1-c_0]\label{c0ODE}\\
 \frac{d A}{d t}&=& \mu(1-c_0)-(\nu c_0+d)A.\label{ODEA}      
 \end{eqnarray}   
 
We first find the uniform steady-state (USS) solutions (fixed points) in (\ref{PDE2}) and then carry out standard  linear stability analysis near the  USSs. From (\ref{c0ODE}) and (\ref{ODEA}), one can see easily that the trivial USS $(c_0,A)=(0,\mu/d)$ always exists, and  other  non-trivial USSs $(c_0^*\neq 0, A^*)$ can also exist. The number of non-trivial USSs depends on the parameter regimes, and the values of $(c_0^*, A^*)$ depend on   $\mu$, $\nu$, $d$, and the parameters in ${\cal P}(A)$. The details of the calculations of the fixed points and linear stability analysis for general feedback function ${\cal P}(A)$ are shown in Appendix IA. The condition for the existence of a physical ($c_0^*>0$) non-trivial USS is rather general, it only requires the existence of a root in (\ref{Astar}) which satisfies (\ref{Astar2})(see Appendix IA ).
The properties of the uniform steady states and their transitions in the system is determined by  the trivial  and non-trivial fixed point(s) and their stabilities.

 For the Hill form feedback function in (\ref{PA}),  as shown in Appendix IB, the  stabilities of the USSs  are determined only by the following four positive parameters: ${\tilde \mu}\equiv\frac{\gamma\mu}{d}$, ${\tilde \nu}\equiv\frac{\nu}{d}$, $m$ and $p$. Here $m$ is a positive real number but not necessarily constrained to be an integer. For $0<p<\tfrac{1}{2}$, only the trivial USS exists and is stable. And for $\tfrac{1}{2}\leqslant p\leqslant 1$,  the stability boundary  of the trivial USS $(0,\mu/d)$ is (unstable if ${\tilde \mu}< {\tilde \mu}_0(p)$)
\begin{equation}
 {\tilde \mu}={\tilde \mu}_0(p)\equiv (2p-1)^{1\over m}\label{mu0p}.
\end{equation}
The non-trivial USSs  fixed point $\gamma A^*$   can be derived from (\ref{Astar}) and is given by the root of
\begin{equation}
F(X)\equiv (1-{\tilde \nu}) X^{m+1}-{2{\tilde \mu}}X^m+(1-{\tilde \nu}+{2p{\tilde \nu}})X -{2{\tilde \mu}(1-p)}=0,\label{FX}
\end{equation}
where $X\equiv \gamma A^*$.
 
 The number of real and positive roots of $X$  depends on the range of values of ${\tilde \mu}$ and can be derived  analytically (details are shown in Appendix IB.2). The critical value, ${\tilde \mu}_t$ at which the number of positive roots changes from 1 to 3 (for ${\tilde \nu} <1$) or 0 to 2 (for ${\tilde \nu} \geqslant 1$) can be obtained from the solution of 
$F(X_t)=0$ and $F'(X_t)=0$, where $X_t$ is the corresponding value of the root at ${\tilde \mu}_t$. ${\tilde \mu}_t$ has two branches and are given by
\begin{equation}
2{\tilde \mu}_t^{\pm}=\frac{m(1-{\tilde \nu})Y_{\pm}^{1+{1\over m}}}{(m-1)Y_{\pm}-1+p},\label{mutpm}
\end{equation} where
\begin{eqnarray}
2(1-{\tilde \nu})Y_{\pm}&=&[mp(1+{\tilde \nu})+(2-3p){\tilde \nu}+p-2]\nonumber
\\& &\pm \sqrt{[mp(1+{\tilde \nu})+(2-3p){\tilde \nu}+p-2]^2+(1-p)(1-{\tilde \nu}+{2p{\tilde \nu}})}.\label{Ypm}
\end{eqnarray}
 
 The number of roots in $X$ (non-trivial fixed points) (see Table \ref{tabroots}) and their stability depend on the regime set by ${\tilde \mu}_t^\pm (p)$ given by (\ref{mutpm}) and the stability  boundary ${\tilde\mu}_0(p)$ also. See Appendix IB for complete calculations.  For physically possible states, both $c_0$ and $A$ have to be real and non-negative, and the phase diagrams in Fig. \ref{phase} summarize regions of stable physical uniform steady states on the ${\tilde \mu} \geqslant 0$ and $0.5\leqslant p \leqslant 1$ plane. The nature of bifurcation and the  phase diagram can be classified into two types according to ${\tilde \nu}<1$ or  ${\tilde \nu}\geqslant 1$ as follows.

\subsubsection{ ${\tilde \nu}<1$}
First consider the slow proliferation case of ${\tilde \nu}<1$, in which the stem cell replication rate is less than the decay rate of the regulating molecules. One can see from Table \ref{tabroots} that there can be three non-trivial positive roots for $X$ for ${\tilde \mu}_t^{+}<{\tilde \mu}<{\tilde \mu}_t^-$, and 1 positive root otherwise. ${\tilde \mu}_t^{+}$ and ${\tilde \mu}_t^{-}$ approach each other as $p$ decreases and there is a threshold $p_t({\tilde \nu})$ (see (\ref{pcnu}) in Appendix IB.2) below which the 3-root regime vanishes. 
Detail examination indicates that at most one (or none) of them is both physical ($X\leqslant {\tilde \mu}$ or $c_0^*\geqslant 0$) and stable, depending on $p$ is greater than or less than some critical value $p_c$, which will be analyzed in details as follows:
The ${\tilde \mu}_t^-$ and ${\tilde \mu}_0$ curves crosses at some `` critical" value of  $p_c$, which can be derived analytically as follows. At  $p_c$, ${\tilde \mu}={\tilde \mu}_0$ and the corresponding root $X$ satisfies 
$F(X)=F'(X)=0$. Therefore, $p_c({\tilde \nu})$ can be calculated simply by requiring 
 $F'(X={\tilde \mu}_0)=0$
  and hence $p_c$ can be derived to  give
\begin{equation}
p_c({\tilde \nu})=\frac{m(1+{\tilde \nu})}{2[m(1+{\tilde \nu})-1]}.\label{pstarnu}
\end{equation}
Therefore, a stable non-trivial USS exists for
\begin{equation}
{\tilde \mu}< \begin{cases}  {\tilde \mu}_t^-(p)  &\mbox{if }p >p_c({\tilde \nu}) \\
{\tilde \mu}_0(p) 
 &\mbox{if }p \leqslant p_c({\tilde \nu}) 
 \end{cases}.\label{muless}
\end{equation}

\begin{figure}[htbp]
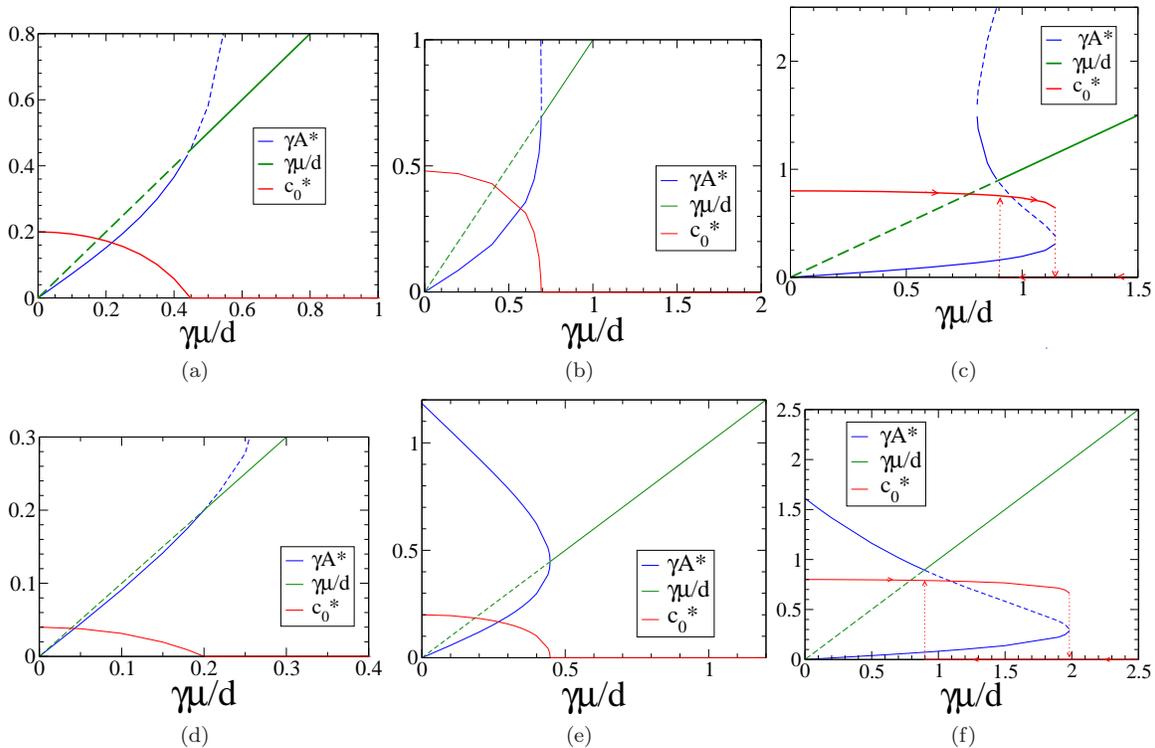

\centering
\subfigure[]{\includegraphics*[width=\figwss]{p_6m_2Ac0.eps}}
\subfigure[]{\includegraphics*[width=\figwss]{nu_5m2p_74.eps}}
\subfigure[]{\includegraphics*[width=\figwss]{p_9m_2Ac0.eps}}
\subfigure[]{\includegraphics*[width=\figwss]{nu2m2p_52.eps}}
\subfigure[]{\includegraphics*[width=\figwss]{nu2m2p_6.eps}}
\subfigure[]{\includegraphics*[width=\figwss]{nu2m2p_9.eps}}
      \caption{The trivial (green curve) and non-trivial (blue curve) fixed points $\gamma A$ as a function of ${\tilde \mu}\equiv \frac{\gamma\mu}{d}$ for  $ m=2$. The solid lines show the stable states and the dashed lines show the unstable states.  The corresponding physical ($c_0\geqslant 0$) and stable $c_0$ states are also shown by the red curve. (a)   ${\tilde \nu}\equiv\frac{\nu}{d}=0.5$. $ p= 0.6$  (b)  ${\tilde \nu}=0.5$. $ p= 0.74$. (c)  ${\tilde \nu}=0.5$. $ p= 0.9$.  The  red vertical dotted lines shows the hysteresis behavior and the bistable phase region lies inbetween them. (d) ${\tilde \nu}=2$. $ p=0.52$.
      (e)  ${\tilde \nu}=2$. $ p=0.6$. (f)  ${\tilde \nu}=2$. $ p=0.9$.}
\label{bifurcation}
\end{figure}

Fig. \ref{bifurcation}a$-$c show the typical bifurcations for $p$ in different regimes. The non-trivial positive roots for $X\equiv \gamma A^*$ are shown as a function of ${\tilde \mu}$ in Fig. \ref{bifurcation} and their stabilities are denoted by solid (stable) and dashed (unstable) lines. In addition, the non-trivial root is physical (i.e. $c_0^* \geqslant 0$) only  for $X\leqslant {\tilde \mu}$. The trivial root $\gamma A={\tilde \mu}$ is also shown (green $y=x$ straight line) whose stability is also denoted by solid (stable) and dashed (unstable) portions.  The corresponding value of $c_0$ for the physical and stable state is also shown (red solid curve).
For $p<p_c({\tilde \nu})$ (see Fig. \ref{bifurcation}a for $p<p_t$ in the single non-trivial root regime and  \ref{bifurcation}b for $p_t<p<p_c$ in the 3-root regime), there is a continuous transition 
at ${\tilde \mu}_0(p)$ from the non-trivial USS to the trivial USS as ${\tilde \mu}$ increases. At  ${\tilde \mu}_0(p)$ , the trivial and non-trivial states exchange their stabilities, signifying a flip bifurcation for the transition between these two USSs (see Fig. \ref{bifurcation}a and b.) On the other hand, there is a bistable regime for  ${\tilde \mu}_t^-({\tilde \nu},p)>{\tilde \mu}>{\tilde \mu}_0(p)$ for  $p>p_c({\tilde \nu})$ in 
which the trivial and non-trivial USSs coexist. There is a first-order transition, characterized by a hystersis loop (indicated by the arrows for the $c_0$  curve) from the non-trivial USS to the trivial USS as ${\tilde \mu}$ increases.

\begin{figure}[htbp]
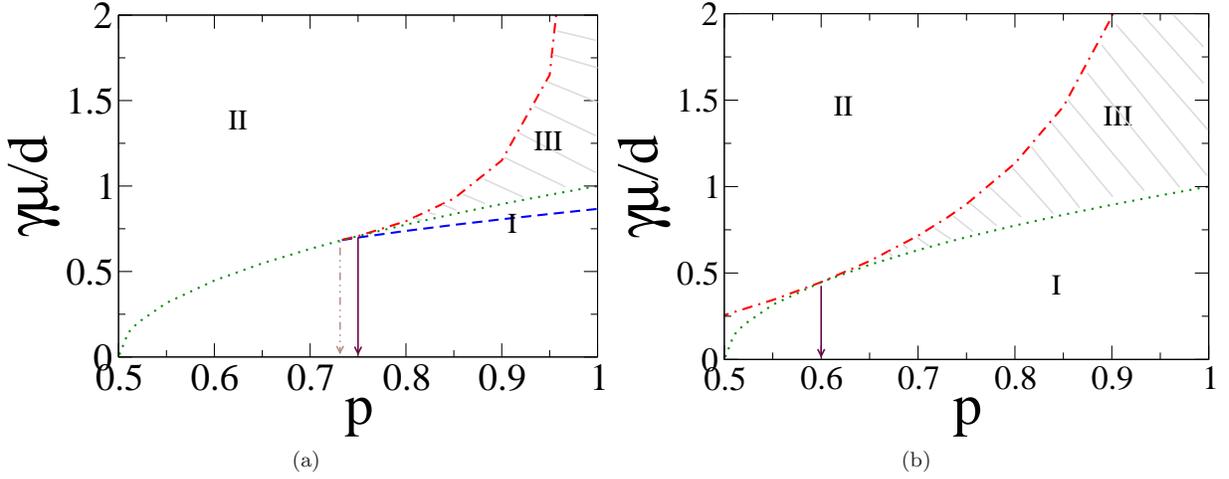

\centering
\subfigure[]{\includegraphics*[width=\figws]{nu_5m_2phase.eps}}
\subfigure[]{\includegraphics*[width=\figws]{nu2m_2phase.eps}}
      \caption{The phase diagram of ${\tilde \mu}\equiv\frac{\gamma\mu}{d}$ vs $p$ with  $ m=2$. The phase boundaries  for the number of roots from equation (\ref{mutpm}), ${\tilde \mu}_t^+$ and ${\tilde \mu}_t^-$ are denoted by blue dashed and red dot-dashed curves  respectively. The  ${\tilde \mu}={\tilde \mu}_0(p)\equiv (2p-1)^{1\over m}$ curve is denoted by a green dotted curve separating the blow-up growth (region I) and final-state growth (region II). The bistable regime (hatched region III) in which the $c_0=0$ (trivial state) and $c_0>0$ (non-trivial state) uniform steady states coexist is marked by the shaded region between the ${\tilde \mu}_t^-$ and ${\tilde \mu}_0$ curves.  (a) ${\tilde \nu}\equiv \frac{\nu}{d}=0.5$.  The critical $p_t(\nu)$ given by (\ref{pcnu}) and $p_c(\nu)$ given by (\ref{pstarnu}) are shown by the vertical dot-dashed arrow and solid arrow respectively. (b) 
Similar phase diagram for            ${\tilde \nu}=2$.}
\label{phase}
 \end{figure}

The properties of the USSs  is summarized in the phase diagram of ${\tilde \mu}$ vs. $p$  shown in Fig. \ref{phase}.
The 3-root regime of the non-trivial state is bounded by the ${\tilde \mu}_t^-$ and ${\tilde \mu}_t^+$ curves which merge together at $p_t$ as shown in Fig. \ref{phase}a. For $p<\tfrac{1}{2}$, only region II exists. The stability boundary for the trivial USS, ${\tilde \mu}_0(p)$, is also shown.
Since the stable trivial USS lies in the  ${\tilde \mu}\geqslant {\tilde \mu}_0(p) $ regime, there is a bistable region  with the coexistence of the  trivial and  non-trivial USS (denoted by the shaded region in Fig. \ref{phase}a). 

\subsubsection{ ${\tilde \nu}\geqslant 1$}
For the rapid proliferating case of ${\tilde \nu}\geqslant 1$,  the stem cell replication rate is faster than the decay rate of the regulating molecules. From Table \ref{tabroots}, one can see that  there can be two positive non-trivial roots $X\equiv \gamma A^*$, but careful examination reveals that only one of them is both physical ($X\leqslant {\tilde \mu}$) and stable for $p>p_c$. For $p\leqslant p_c$, the two non-trivial positive roots are both unphysical ($X> {\tilde \mu}$). Thus the non-trivial stable USS  again lies in the region given by (\ref{muless}).

Fig. \ref{bifurcation}d$-$f show the typical bifurcations for different regimes of $p$. For $p\leqslant p_c({\tilde \nu})$ (see Fig. \ref{bifurcation}d for $p<p_c$  and   \ref{bifurcation}e for $p=p_c$, both in the 2-root regime), there is a continuous transition 
at ${\tilde \mu}_0(p)$ from the non-trivial USS to the trivial USS as ${\tilde \mu}$ increases. For $p<p_c$, the trivial and non-trivial states exchange their stabilities, signifying a flip bifurcation  at ${\tilde \mu}_0(p)$   (see Fig. \ref{bifurcation}d).  
For  $p>p_c({\tilde \nu})$,   the trivial and non-trivial USSs coexist associated with a hystersis loop in the ${\tilde \mu}_0(p)<{\tilde \mu}<{\tilde \mu}_t^- $ regime as shown in  Fig. \ref{bifurcation}f.
The phase diagram for the ${\tilde \nu}\geqslant 1$ case is shown in Fig. \ref{phase}b.
Since the stable trivial  USS  lies in the region 
${\tilde \mu}\geqslant {\tilde \mu}_0(p) $, there is a coexisting bistable region for these two USSs for $p >p_c({\tilde \nu})$, as marked by the hatched region.
\begin{figure}[htbp]
\centering
\subfigure[]{\includegraphics*[width=\figwss]{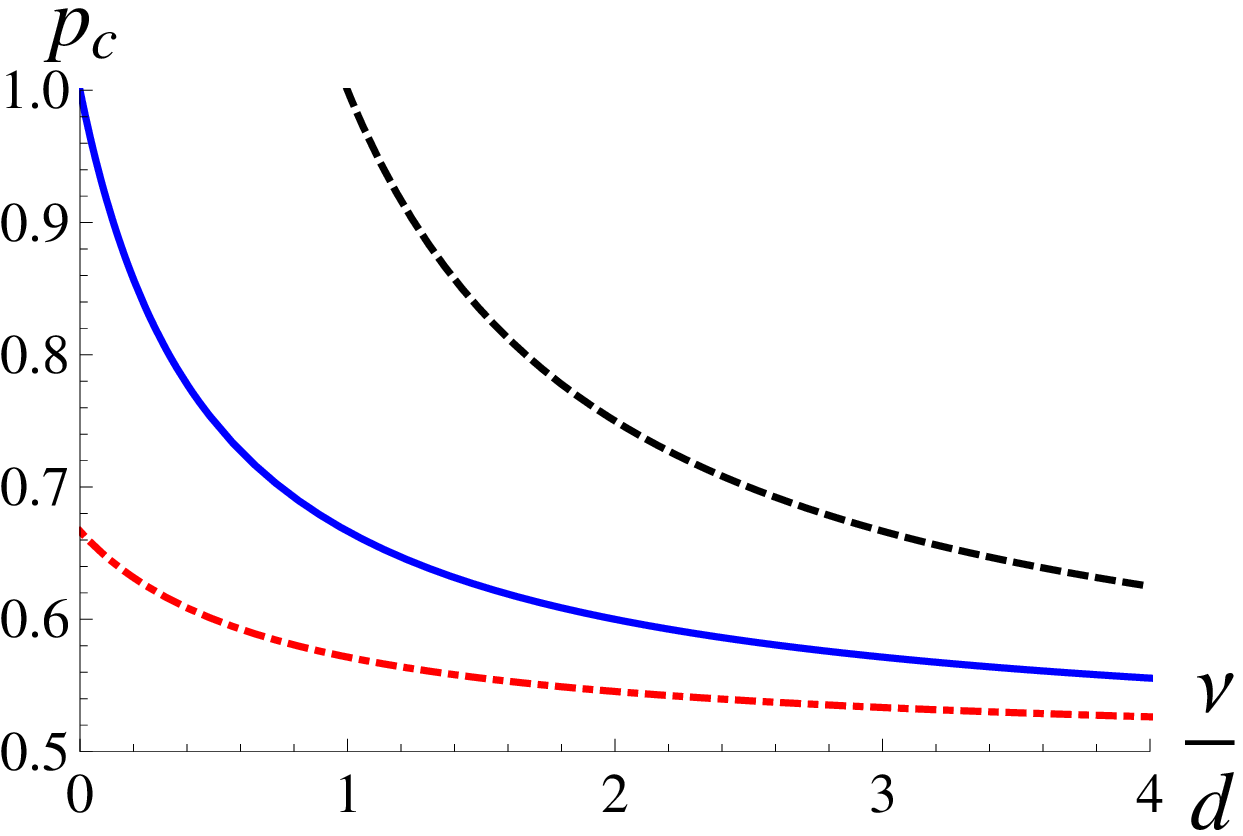}}
\subfigure[]{\includegraphics*[width=\figwss]{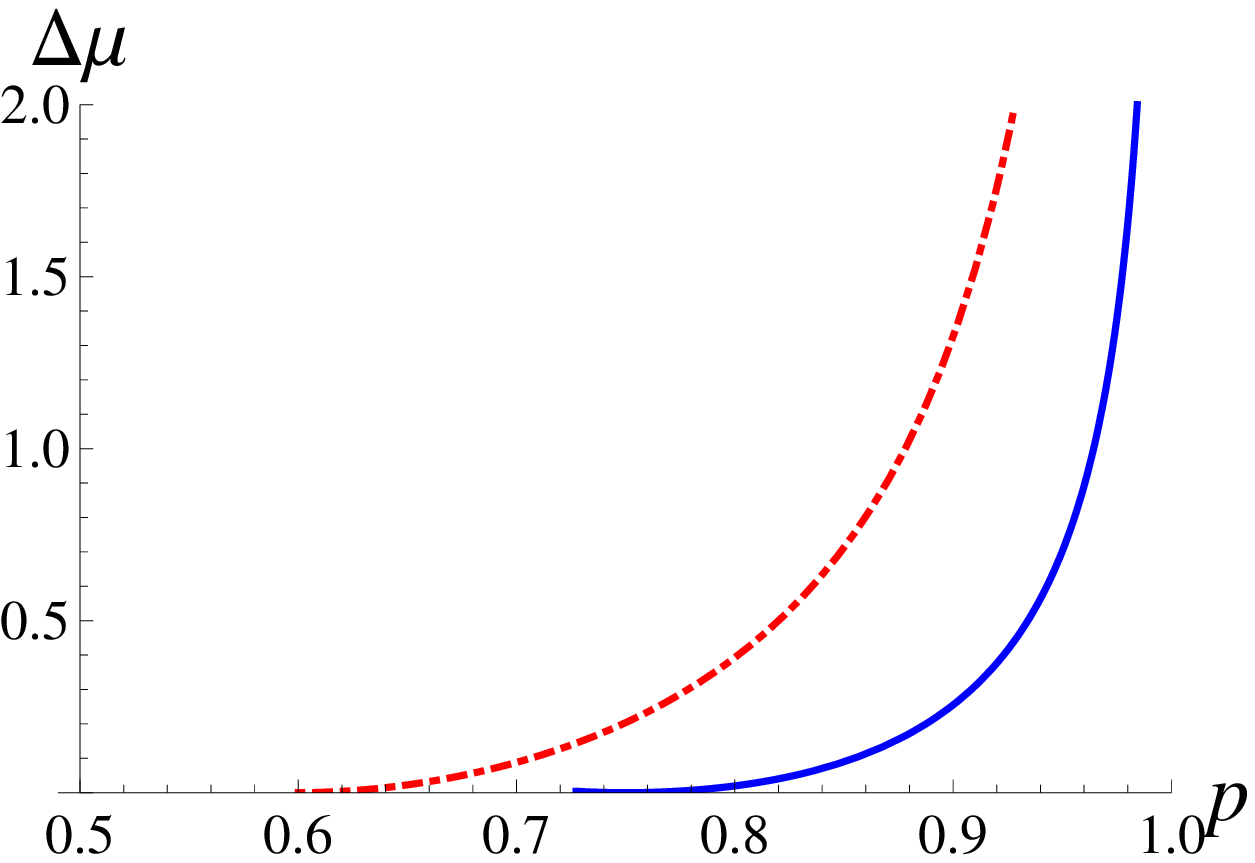}}
\subfigure[]{\includegraphics*[width=\figwss]{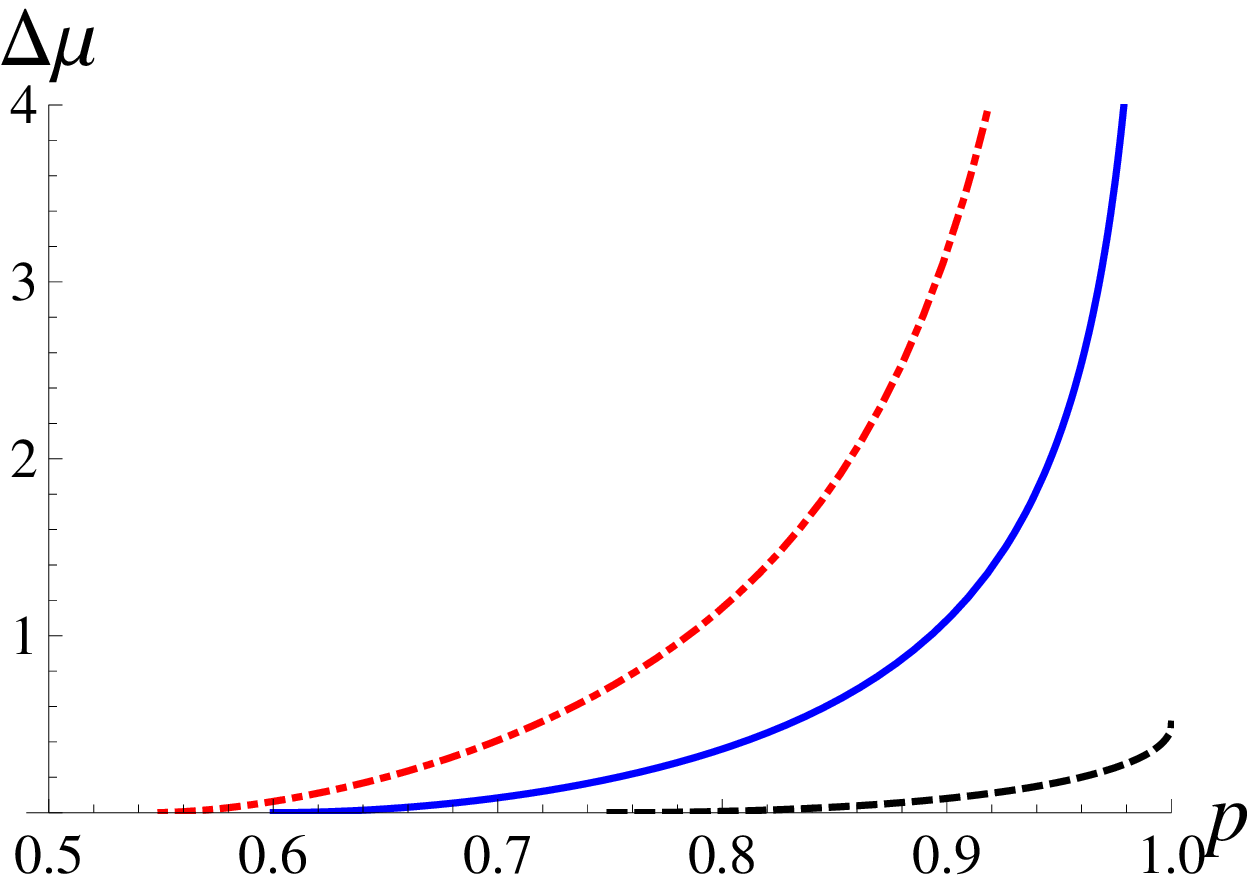}}
      \caption{(a) The critical value at which the coexistence regime vanishes, $p_c$ (given by (\ref{pstarnu}) as a function of $ {\tilde \nu}$ for different values of the Hill coefficient, $m=1$ (dashed black), $m=2$ (solid blue) and $m=4$(dot-dashed red).  (b) The size of the co-existence regime $\Delta{\tilde \mu}\equiv {\tilde \mu}_t^--{\tilde \mu}_0$ (given by (\ref{mutpm}) and (\ref{mu0p}))   as a function of $p$ for different values of $m$ for $ {\tilde \nu}=0.5$ and (c) for $ {\tilde \nu}=2$.}
\label{pcDmu}
 \end{figure}

The critical value at which the coexistence regime vanishes, $p_c$, is displayed as a function of $ {\tilde \nu}$ for different values of the Hill coefficient. $p_c$ decreases monotonically with ${\tilde \nu}$ indicating the coexistence regime increases with ${\tilde \nu}$. $p_c$ is  also smaller for larger values of $m$ suggesting the coexistence regime is larger for larger $m$.
The size of the co-existence regime can further be characterized by $\Delta{\tilde \mu}\equiv {\tilde \mu}_t^--{\tilde \mu}_0$. Fig. \ref{pcDmu}b and c show  $\Delta{\tilde \mu}$ as a function of $p$ for different values of $m$ for $ {\tilde \nu}=0.5$ and 2 respectively, indicating that the bistable regime  increases with $p$ and is larger for larger values of  $m$. Notice that for ${\tilde \nu}<1$, there is no bi-stable regime for $m=1$.

Summarizing this section briefly in physical terms: the trivial USS and non-trivial USS correspond to the final-state and blow-up state respectively. The existence of the latter can be analytical  tracked by the physical root of  $F(X)$ in (\ref{FX}). The stability of these two states can also be calculated analytically to determine their respective stabilities and the condition of the coexistence of the final-state and blow-up state. The full phase diagram of the model and the corresponding phase boundaries for the stable regimes can be calculated analytically in terms of the parameters of the model.

\section{Analytic Results: Tissue Size Dynamics}
Denoting the leading edge of the issue at time $t$ by $L_m(t)$, and consider initially the SC and TD cells occupy uniformly in $0\leqslant z \leqslant L_m(0)$, where $L_m(0)$ is the initial tissue size. The cells will grow and advect with speed $V$ and hence  the tissue size, $L_m(t)$ will increase with time, whose dynamics can be obtained from our model equations.  To determine the evolution dynamics of the tissue size, it should be noticed that even though the non-trivial USS is stable theoretically,  there is a  sharp cell density gradient in the leading edge ($z=L_{m}$) of the tissue in practice.  Such a sharp gradient in $c_0$ can destablize the system and lead to the  blow-up growth of the tissue. Such a scenario can be understood theoretically from our model.
The growth rate of the leading edge is given by the advection speed, thus we have from (\ref{Vz2})
\begin{equation}
\frac{dL_{m}}{dt}= V(z=L_{m})=\nu\int_0^{L_m} c_0(x,t) dx.\label{LmODE}
\end{equation}
 The tissue growth acceleration can also be calculated by differentiating (\ref{LmODE}) to be
\begin{equation}
\frac{d^2 L_m}{dt^2}=\nu^2\int_0^{L_m} dz c_0(z,t)[2{\cal P}(A(z,t))-1]\label{Lmacc}
\end{equation}
  and thus  it is possible to have an early stage of accelerated growth and then
  slow down to the final tissue size, typical of a realistic ``S-shape" biological growth curve\cite{growthcurve}.
  
   Since we are mostly interested in the dynamics of the  tissue size, rather on the details of the spatial density profiles of the cells or regulatory molecules, we can further approximate the spatial profiles to be step-functions and proceed for further analytic results. As can be seen in the numerical solutions in Sec. IV, the step-function approximation rather good except very near the leading edge. Under the step profile approximation, we have $c_0(z,t)=c_0(t)$ for $0< z \leqslant L_m(t)$ and vanishes otherwise. Since $A$  rapidly equilibrated and from (\ref{PDEA}), $A$ is also a step-profile with magnitude $A=A(t)\equiv \frac{\mu (1-c_0(t))}{d(1+{\tilde \nu}c_0(t))}$ in $0< z \leqslant L_m(t)$. Using (\ref{Lmacc}) and take the initial stem cell profile to be a step function of height $c_0(0)$ and size $L_m(0)$, one obtains the equation of motion for $L_m$ as
   \begin{eqnarray}
   \frac{d^2L_m}{dt^2}&=&\nu^2c_0(t)\left[ 2{\cal P}(A(t))-1\right]L_m \label{Lmacc2}\\
   \frac{dL_m}{dt}|_{t=0}&=& \nu c_0(0)L_m(0),\qquad A(t)\equiv \frac{\mu (1-c_0(t))}{d(1+{\tilde \nu}c_0(t))}.
   \end{eqnarray}
  $L_m(t)$ can be solved together with the equation of motion of $c_0(t)$ which is given in (\ref{c0ODE}).
  
  \subsection{Tissue growth Time scales}
  We first analyze in the tissue size dynamical evolution to the final-state, and one can see from (\ref{c0ODE}) that $c_0(t)$ is always decreasing and eventually approaches to zero. As shown in  Appendix IA, the saturation rate of final-state growth tissue is given by the rate of $c_0$ approaching the trivial USS fixed point, with the saturation time scale
  \begin{equation}
\tau_s^{-1}=\nu[1-2{\cal P}(\tfrac{\mu}{d})].\label{taus}
\end{equation}

   Moreover, the tissue acceleration can be positive or negative and one can derive the condition for the tissue dynamics with a S-shape growth curve as follows, even without the explicit solution of $L_m(t)$.
  The S-shape growth curve is signified by an early acceleration and late time deceleration as it approaches the final-state. There is an inflexion point at $t=\tau_{sw}$ at which the grow acceleration switches from positive to negative,  i.e. $\frac{d^2L_m}{dt^2}|_{t=\tau_{sw}}=0$. From (\ref{Lmacc2}), one can solve for the corresponding $c_0(\tau_{sw})\equiv c_0^{sw}$ to be
    \begin{equation}
  c_0^{sw}= \frac{\mu-d {\cal P}^{-1}(\tfrac{1}{2})}{\mu+\nu{\cal P}^{-1}(\tfrac{1}{2})}\label{c0sw}
  \end{equation}
Since $c_0(t)$ is a decreases with $t$, therefore the inflexion point occurs only if $c_0(0)> c_0^{sw}$, i.e. the initial stem-cell profile cannot be too small to have an early acceleration growth.

Now for the case of blow-up dynamics, one can also estimate the explosive growth time scale as follows. For $z<<L_{m}$, the  stable uniform steady-state non-trivial fixed point dominate and the  cell density $c_0\simeq c_0^*$. But for $z\approx L_{m}$, the large negative gradient $\partial_zc_0$  dominates  over the first term in (\ref{PDE3}) and destablizes the leading edge to give rise to blow-up growth in the tissue size.
The   tissue size increases exponentially with a time-scale $\tau$, which can be estimated as follows. The growth rate of the leading edge in this case can be estimated by approximating the SC profile with a step function of height $c_0^*$ and size $L_{m}$, thus we have from (\ref{LmODE})
\begin{equation}
\frac{dL_{m}}{dt}\simeq\nu c_0^* L_{m}.\label{LmODEblowup}
\end{equation}
    It then follows that the tissue size grows exponentially with a characteristic time-scale of $\tau$ given by
\begin{equation} \tau=\frac{1}{\nu c_0^*},\label{tau}
  \end{equation}
which can be checked against the values obtained from the fitting of the numerical solutions.
It should be noted that if the initial SC concentration is far from the  USS value $c_0^*$, then the system will need a few cycle times to be attracted near the value of $c_0^*$ and then the tissue size will grow exponentially with the time-scale given by (\ref{tau}).


\subsection{Tissue size of the final-state}
Here we derive an approximate formula for the ultimate tissue size for the final-state growth, $L_m(t\to\infty)$. We shall focus on the case in which the final-state is the only stable state (i.e. $p<p_c({\tilde \nu})$) characterized by the trivial fixed point. Since for large $t$, $c_0$ decays to small values, expanding (\ref{c0ODE}) to leading order in $c_0$, one gets
\begin{equation}
 \frac{dc_0}{dt}=-\frac{c_0}{\tau_s} +{\cal O}(c_0^2)
\end{equation}
where $\tau_s$ is given by (\ref{taus}). Hence
\begin{equation}
c_0(t)\simeq \alpha e^{-\tfrac{t}{\tau_s}}\label{c0exp}
\end{equation}
for some constant $\alpha$ to be determined. Note that the decay time scale $\tau_s$ in (\ref{c0exp}) agrees with that in (\ref{taus}).
With the same step-profile approximation as in previous subsection, $L_m(t)$ is still given by (\ref{Lmacc}), but for large $t$, it has to satisfy the boundary condition of 
$\frac{dL_m}{dt}|_{t=\infty}=0$. For large $t$ and hence small $c_0$, (\ref{Lmacc}) becomes
 \begin{eqnarray}
   \frac{d^2L_m}{dt^2}&\simeq&\nu^2c_0(t)\left[ 2{\cal P}(\tfrac{\mu}{d})-1\right]L_m\\
   &\simeq& -K^2 e^{-\tfrac{t}{\tau_s}}L_m,\qquad K^2\equiv \tfrac{\nu}{\tau_s}\alpha,\label{LmbigtODE}
\end{eqnarray}    
where (\ref{c0exp}) was used to obtain (\ref{LmbigtODE}). The solution of (\ref{LmbigtODE}) with $\frac{dL_m}{dt}|_{t=\infty}=0$ is
\begin{equation}
L_m(t)=L_m(\infty) J_0(2K\tau_s e^{-\tfrac{t}{2\tau_s}}).  \label{LmtBessel}
\end{equation}
Notice that the Bessel function of the first kind has the expansion $J_0(x)\simeq 1 -\tfrac{x^2}{4} +{\cal O}{x^4}$for small $x$, and hence (\ref{Lmbigt}) agrees with the saturation approach to $L_m(\infty)$ with the time scale of $\tau_s$ discussed earlier and will be verified by the numerical solution in next section.

The constants $\alpha$ and $L_m(\infty)$ can  be  estimated by matching their corresponding values at some (earlier) fixed time, say $t=f\tau_s$ (for some constant fraction $f$), to the extrapolated values from the initial slopes of $c_0(t)$ and $L_m(t)$. After some algebra, one finally gets
\begin{eqnarray}
L_m(\infty)\simeq \frac{1+f\nu\tau_s c_0(0)}{J_0\left(2\sqrt{\nu\tau_s c_0(0)\{1+f\nu\tau_s [2{\cal P}(\frac{\mu (1-c_0(0))}{d(1+{\tilde \nu}c_0(0))})-1-c_0(0)]\} }\right)}  L_m(0),\label{Lmbigt}
\end{eqnarray}
for an initial SC step profile of height $c_0$ and size $L_m(0)$. $f\simeq 0.5$ will be shown to be a reasonable choice in practice.

Summarizing this section briefly: the equation of motion governing the tissue size growth dynamics  is derived and can be solved analytically to obtain  the precise time-dependence, $L_m(t)$ for both the blow-up and final-states. Analytic expression for $L_m(t)$ would be very useful to implement appropriate external control (as illustrated in Sec. VI) or designing upstream regulatory pathways in a timely manner.

\section{Numerical Solutions}

The model equations (\ref{PDE2}) in one spatial dimension can be numerically solved to investigate in details the dynamics of the tissue growth. The numerical results can provide valuable quantitative information on  the time evolution of the tissue size and cell density profiles. Since the regulatory feedback molecules diffuse with a time scale much faster than that of the cell growth dynamics, one can exploit this separation of time scale to solve for the quasi-static spatial distribution of $A(z)$ first and then obtain the cell/tissue dynamics. The details of the numerical method is given in Appendix II.  

\subsection{Blow-up state, final-state and their coexistence}
\begin{figure}[h]
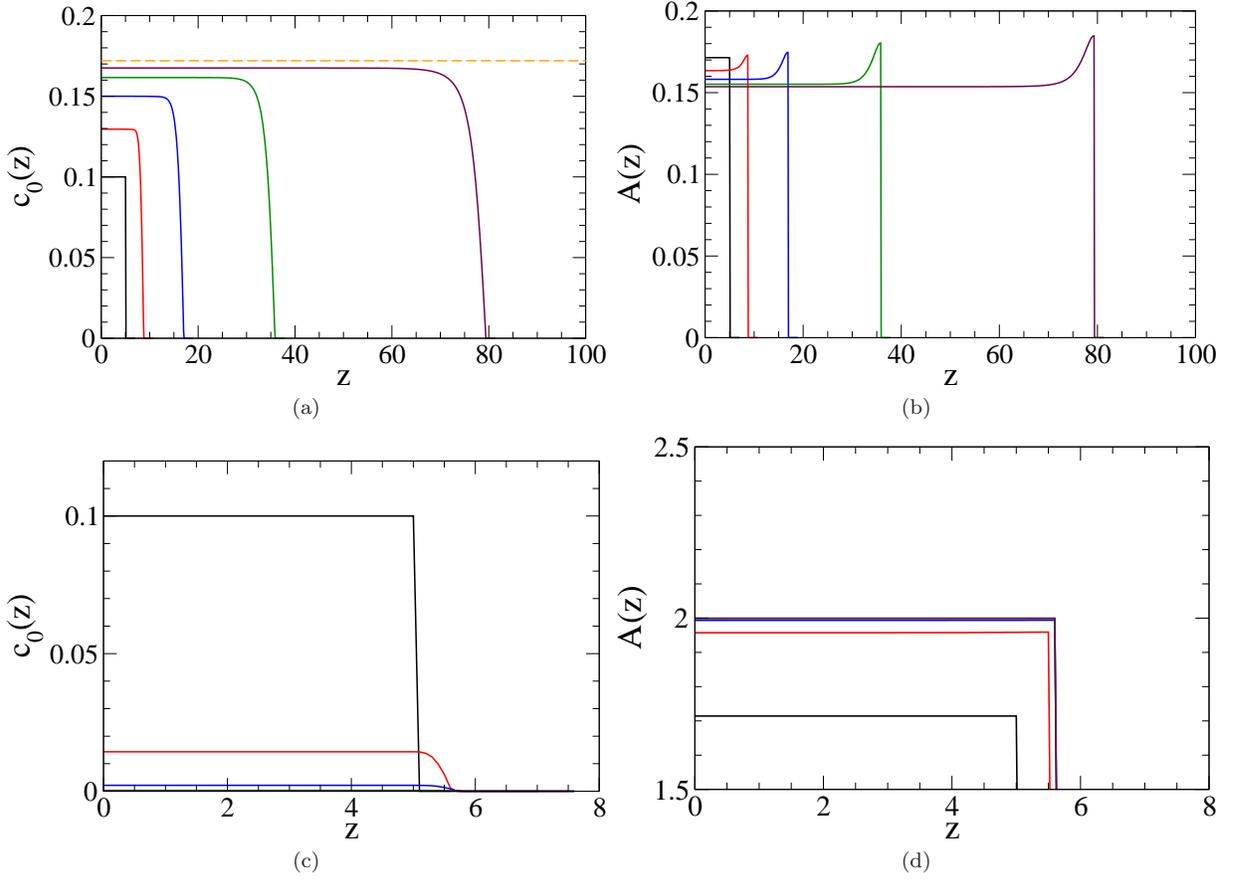

\subfigure[]{\includegraphics*[width=\figws]{c0zmu_2nu_5p_6.eps}}
 \subfigure[]{\includegraphics*[width=\figws]{Azmu_2nu_5p_6.eps}}
\subfigure[]{\includegraphics*[width=\figws]{c0zmu2nu_5p_6.eps}}
 \subfigure[]{\includegraphics*[width=\figws]{Azmu2nu_5p_6.eps}}
      \caption{Numerical solution of the lineage model with $m=2$, ${\tilde \nu} = 0.5$ and $p= 0.6$ for  ${\tilde \mu} = 0.2$ inside region I of the phase diagram in Fig. \ref{phase}a  corresponding to the case of blow-up tissue growth.  Time evolution of the (a) SC distributions $c_0(z)$, (b) feedback molecule concentrations $A(z)$. The non-trivial USS value $c_0^*$ is marked by the horizontal dashed line in (a). 
 (c) and (d) are similar to (a) and (b) for ${\tilde \mu} = 2$ inside region II of the phase diagram in Fig. \ref{phase}a  corresponding to the case of saturated tissue growth to the final-state. Time and space are in  units of $1/d$ and $\sqrt{D/d}$ respectively. Distribution curves  in (a), (b) and (c), (d) are separated by a time of 10 and 5 units respectively. } \label{growth1}\end{figure}

\begin{figure}[h]
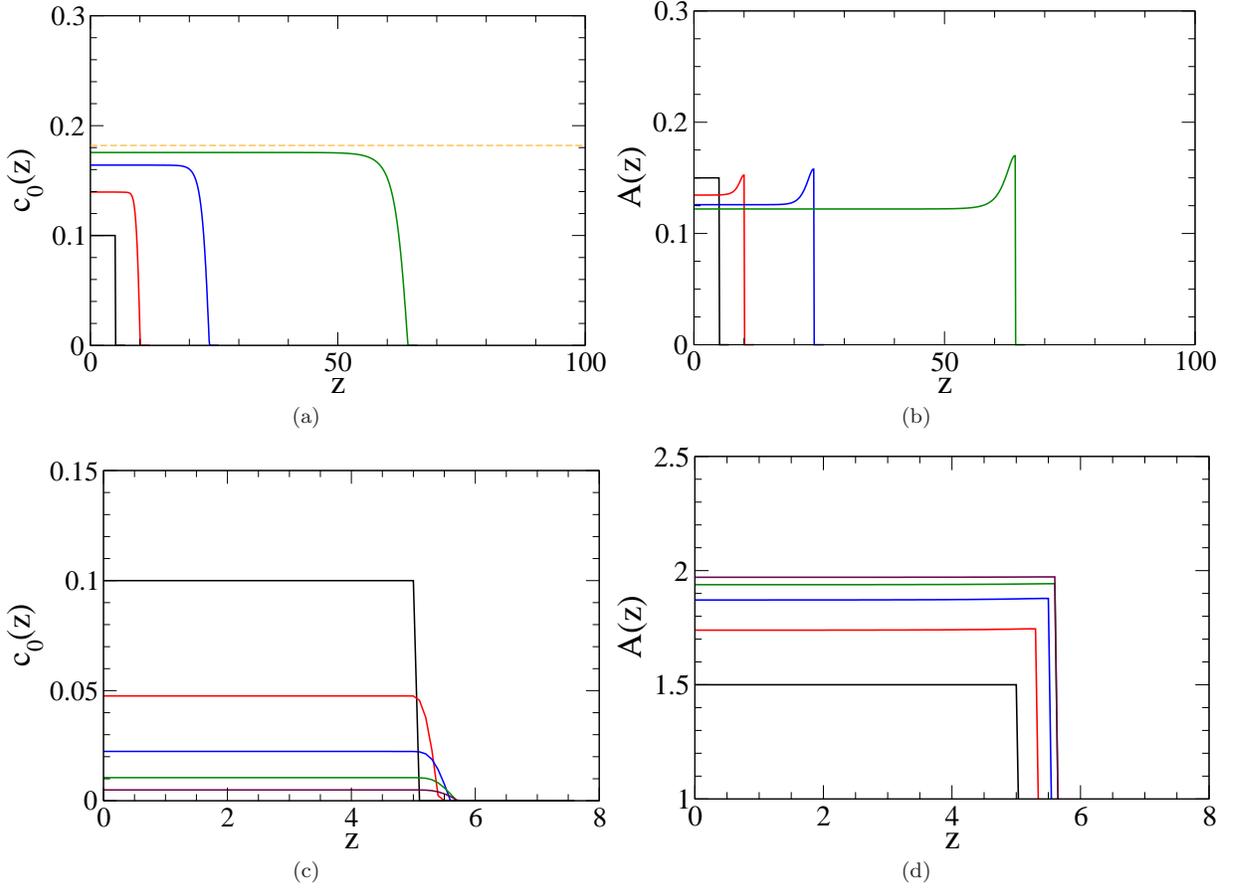

\subfigure[]{\includegraphics*[width=\figws]{c0zmu_2nu2p_6.eps}}
 \subfigure[]{\includegraphics*[width=\figws]{Azmu_2nu2p_6.eps}}
\subfigure[]{\includegraphics*[width=\figws]{c0zmu2nu2p_6.eps}}
 \subfigure[]{\includegraphics*[width=\figws]{Azmu2nu2p_6.eps}}
      \caption{Numerical solution of the lineage model with $m=2$, ${\tilde \nu} = 2$ and $p= 0.6$ for  ${\tilde \mu} = 0.2$ inside region I of the phase diagram in Fig. \ref{phase}b  corresponding to the case of blow-up tissue growth.  Time evolution of the (a) SC distributions $c_0(z)$, (b) feedback molecule concentrations $A(z)$.
 (c) and (d) are similar to (a) and (b) for ${\tilde \mu} = 2$ inside region II of the phase diagram in Fig. \ref{phase}a  corresponding to the case of saturated tissue growth to the final-state. Time and space are in  units of $1/d$ and $\sqrt{D/d}$ respectively. Distribution curves  in (a), (b) and (c), (d) are separated by a time of 3 and 0.5 units respectively. } \label{growth2}\end{figure}

The time evolution of the profiles of $c_0(z)$ and $A(z)$ for $m=2$, ${\tilde \nu} = 0.5$ and $p= 0.6$ for  ${\tilde \mu} = 0.2$ 
 and  ${\tilde \mu} = 2$ are shown in Fig. \ref{growth1}  (region I and II respectively in the phase diagram Fig. \ref{phase}a). As predicted by the analytic result of the phase diagram, region I corresponds to the blow-up growth case as the profile of $c_0(z)$ expands rapidly in space and at the same time the value of $c_0$ also grows and approaches the theoretical non-trivial fixed point value $c_0^*$ (shown by the horizontal dashed line in Fig. \ref{growth1}a).
 The corresponding tissue size growing dynamics is shown in Fig.\ref{Lmvst}b. The dynamical behavior of the blow-up state can be understood qualitatively in terms of the phase space flow as depicted in Fig. \ref{Ac0}b where the system is attracted to the only stable (non-trivial) fixed point $(c_0*>0, A^*)$. And for the dynamics in region II,  the profile of $c_0(z)$ increases very slowly in space and at the same time  $c_0$  approaches to zero (trivial fixed point) and the growth stops as $c_0$ becomes extinct, characterizing the behavior of a final-state.  The corresponding tissue growth dynamics is shown in Fig.\ref{Lmvst}a. Again the asymptotic dynamics is governed by the flow towards the stable (trivial) fixed point $(0,\mu/d)$ as qualitatively shown in Fig. \ref{Ac0}a.
For the case of ${\tilde \nu} =2$, the time evolution of the profiles of $c_0(z)$ and $A(z)$ are shown in Fig. \ref{growth2} for  ${\tilde \mu} = 0.2$ 
 and  ${\tilde \mu} = 2$ corresponding to region I and II respectively in the phase diagram Fig. \ref{phase}b. Compared with Fig. \ref{growth1} for ${\tilde \nu} =0.5$, the growth dynamics is similar qualitatively but is about 4 times faster corresponding to a 4 times larger ${\tilde \nu}$.

\begin{figure}[h]
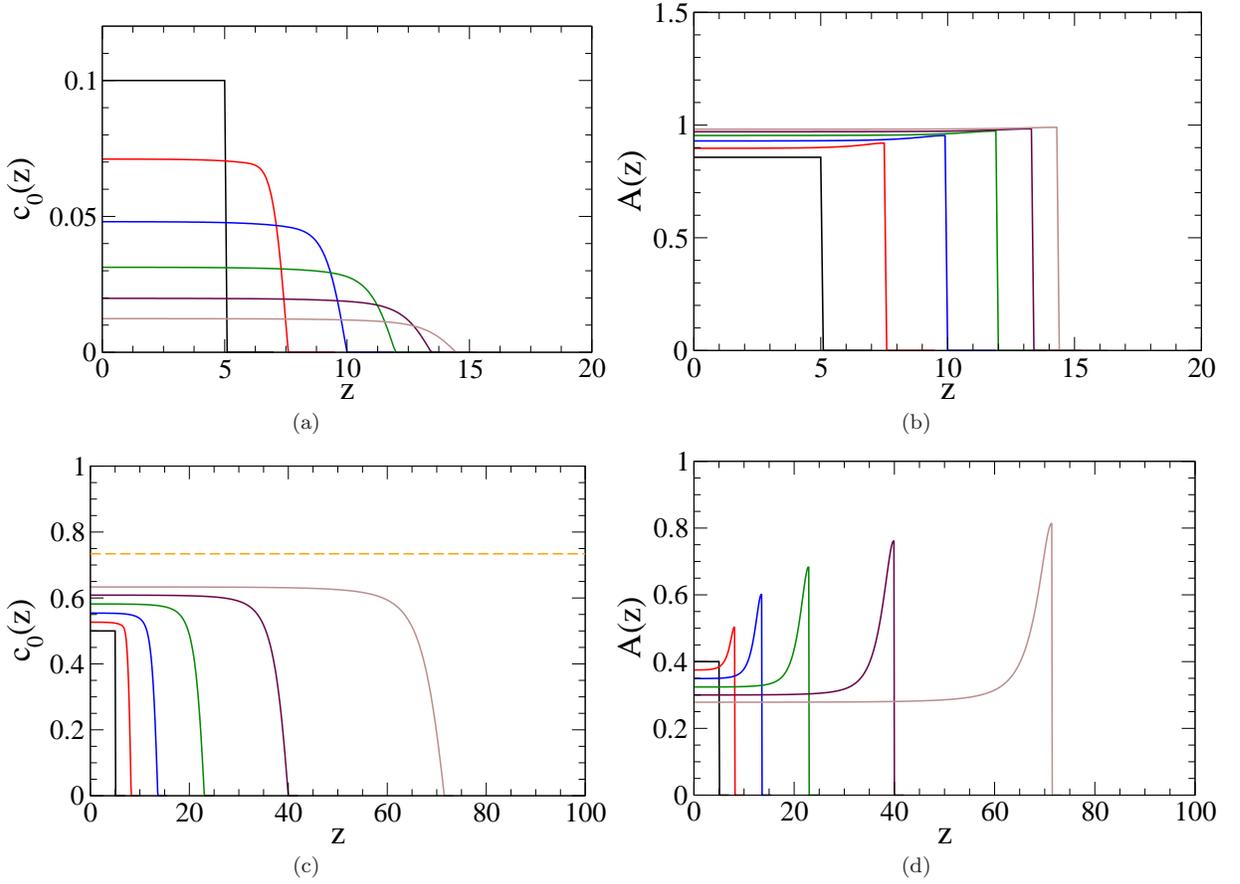

 \subfigure[]{\includegraphics*[width=\figws]{c0zmu1nu_5p_9.eps}}
 \subfigure[]{\includegraphics*[width=\figws]{Azmu1nu_5p_9.eps}}
  \subfigure[]{\includegraphics*[width=\figws]{c0zmu1nu_5p_9b.eps}}
 \subfigure[]{\includegraphics*[width=\figws]{Azmu1nu_5p_9b.eps}}
      \caption{Numerical solution of the lineage model with $m=2$, ${\tilde \nu} = 0.5$ and $p= 0.9$ for  ${\tilde \mu} = 1$ inside the bistable region of the phase diagram in Fig. \ref{phase}a.  Time evolution of the (a) SC distributions $c_0(z)$, (b) feedback molecule concentrations $A(z)$, for initial concentration of $c_0=0.1$.  (c) and  
 (d) are similar to (a) -(b) but for initial concentration of $c_0=0.5$.
 Distribution curves in (a),(b) and (c),(d) are separated by  time intervals of 10 and 2 respectively.  Time and space are in  units of $1/d$ and $\sqrt{D/d}$ respectively.
  } \label{growth3}\end{figure}

\begin{figure}[h]
     \subfigure[]{\includegraphics*[width=\figwss]{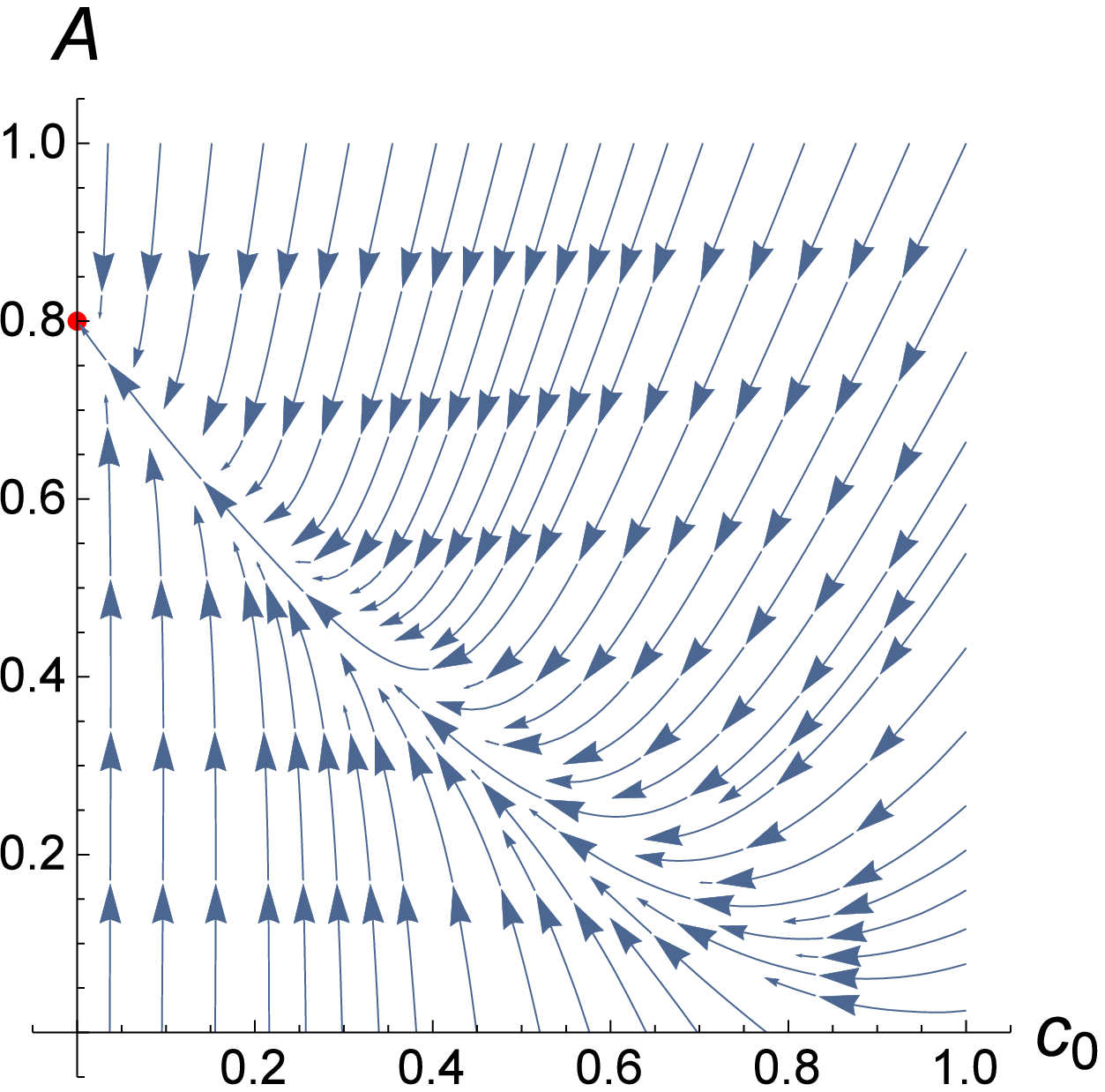}}
      \subfigure[]{\includegraphics*[width=\figwss]{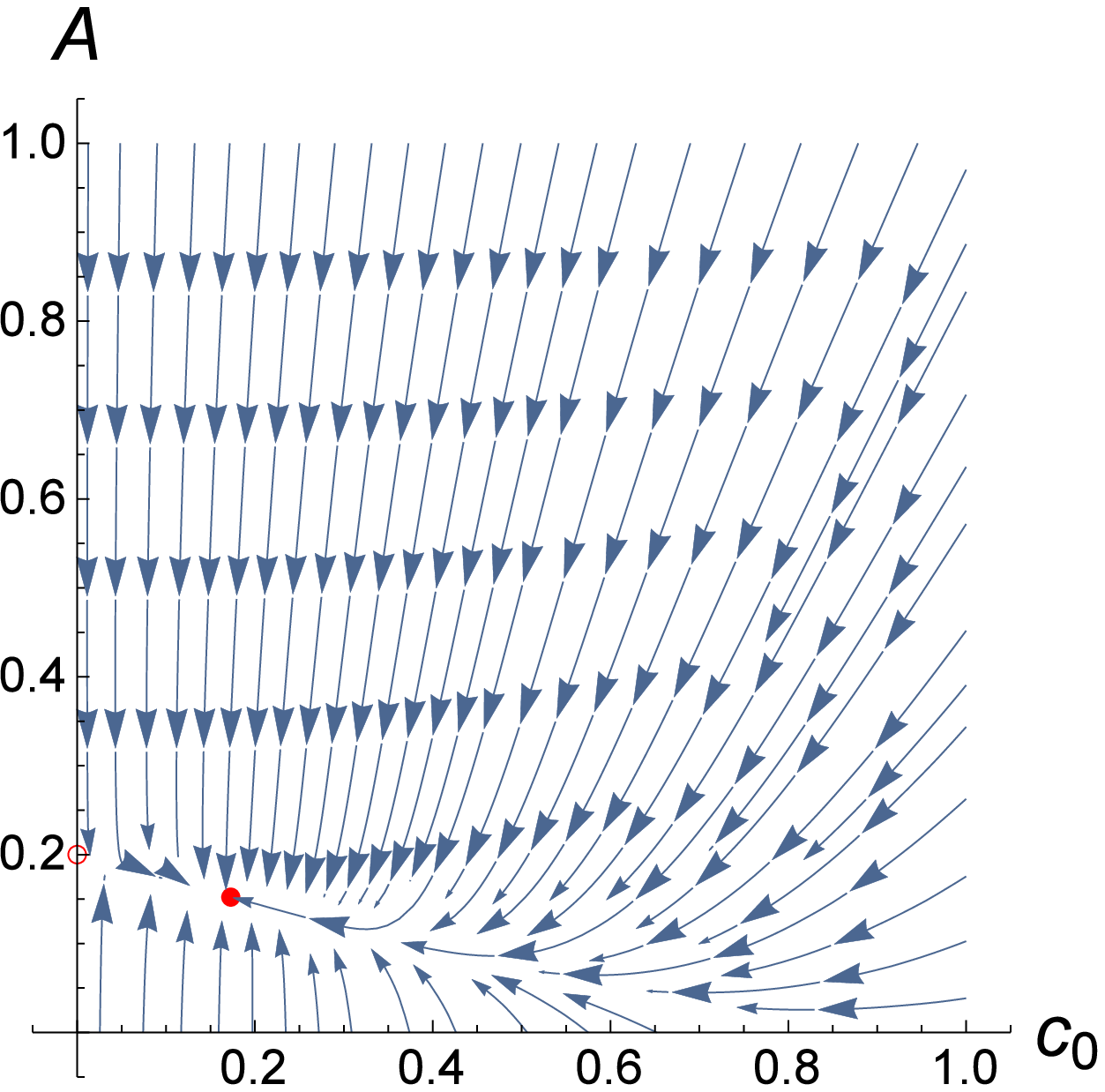}}
      \subfigure[]{\includegraphics*[width=\figwss]{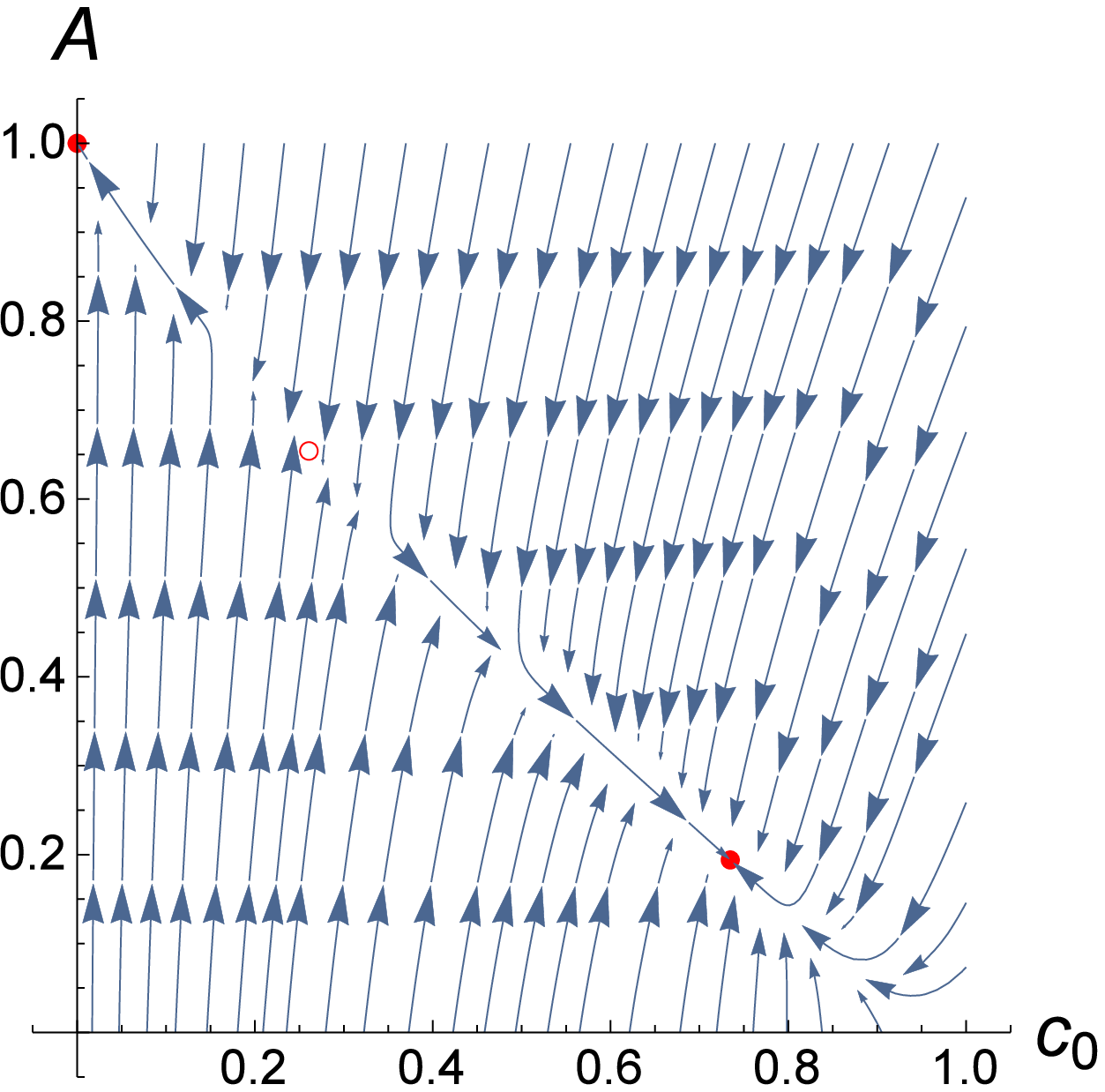}}
  \caption{ The $A$ vs. $c_0$ phase plane portrait obtained from (\ref{c0ODE}) and (\ref{ODEA}) showing (a) the case of final-state with $p=0.6$, ${\tilde \mu}=0.8$, (b)   the case of blow-up state with $p=0.6$, ${\tilde \mu}=0.2$, (c) in the bistable regime with $p=0.9$, ${\tilde \mu}=1$.  In all cases, ${\tilde \nu}=0.5$ and $\gamma=1$. The stable and unstable fixed points are denoted by a filled  and open circles respectively.
  } \label{Ac0}\end{figure}

 We now turn to the more interesting bi-stable situation as  predicted in previous section. From the phase diagrams in Fig. \ref{phase}, we compute the numerical solutions  for the case of ${\tilde \mu} = 1$, $p=0.9$ and ${\tilde \nu} = 0.5$. The  time evolution of the profiles of $c_0(z)$ and $A(z)$ are shown in Fig. \ref{growth3} for two different initial values of the step function profiles of $c_0=0.1$ and $c_0=0.5$ (both are of the same initial spatial extend of 5).  The corresponding tissue size growing dynamics is shown in Fig.\ref{Lmt}a. The fates of the two different initial profiles are totally different and are governed by  the trivial and non-trivial fixed points corresponding to final-state and blow-up growth respectively. As shown in Fig. \ref{Ac0}c for this bistable regime, there are two stable fixed points $(c_0,A)=(0,1)$ and (0.735,0.194) separated by an unstable fixed point (0.261,0.654).  The dynamics in the bistable region can be understood in terms of the flow in the phase plane  showing the two stable fixed points and an unstable one separating their basins of attraction. The ultimate fate of the system depends on the initial SC density that lies in the corresponding attractive basin of one of the two stable fixed point.
 As shown in  Fig.  \ref{growth3}a and Fig.  \ref{growth3}b,  the initial profile with $c_0=0.1$ is close to the trivial fixed point and  the subsequent dynamics shows the attraction towards the final-state. On the other hand, the initial profile with $c_0=0.5$ lies in the basin of attraction of the non-trivial stable fixed point and  the  dynamics evolves towards this fixed point (see  Fig.  \ref{growth3}c and the horizontal dashed line) resulting in blow-up growth.
 The dynamics of the system is very sensitive near the unstable fixed point, even small perturbations can alter the fate of the system. 
 
\subsection{Tissue size Dynamics and Different growth modes}
\begin{figure}[htbp]
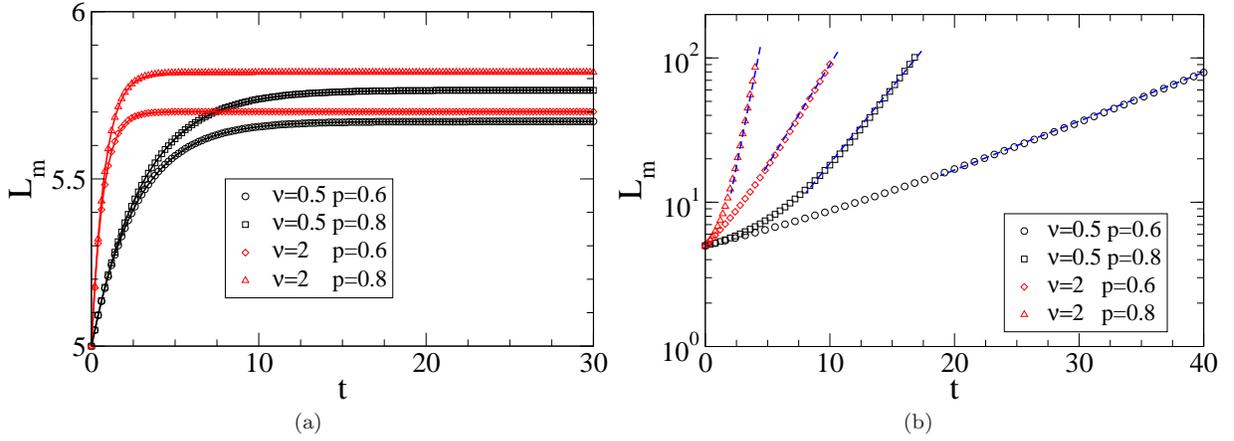

\centering
\subfigure[]{\includegraphics*[width=\figws]{Lmvstmu2m2.eps}}
\subfigure[]{\includegraphics*[width=\figws]{Lmvstmu_2m2.eps}}
      \caption{Numerical results on the tissue size dynamics of the lineage model with $m=2$ for (a) ${\tilde \mu}=2$ corresponding to the case of the tissue size is governed by the final-state for ${\tilde \nu}=0.5$ and 2. The solid curves are fitting of results of the form $L_m(t)=a_0+ a_1 e^{-\frac{t}{\tau_s}}$. (b) Semi-log plot of the tissue size as a function of time for  $\mu=0.2$ corresponds to the case of the blow-up growth. The straight lines in the large time regimes are  fitting using an exponential form from which the exponential growth time-scales $\tau$ are obtained.   Time and $L_m$ are in  units of $1/d$ and $\sqrt{D/d}$ respectively.}
\label{Lmvst}
\end{figure}

 The tissue size, which is an experimentally convenient observable, is also calculated from the numerical solution. 
 Fig. \ref{Lmvst}a  displays the saturation dynamics of the tissue size to the final-state.  
 On the other hand, the tissue size grows exponentially fast for the case of blow-up growth as shown in Fig. \ref{Lmvst}b .

\begin{figure}[h]
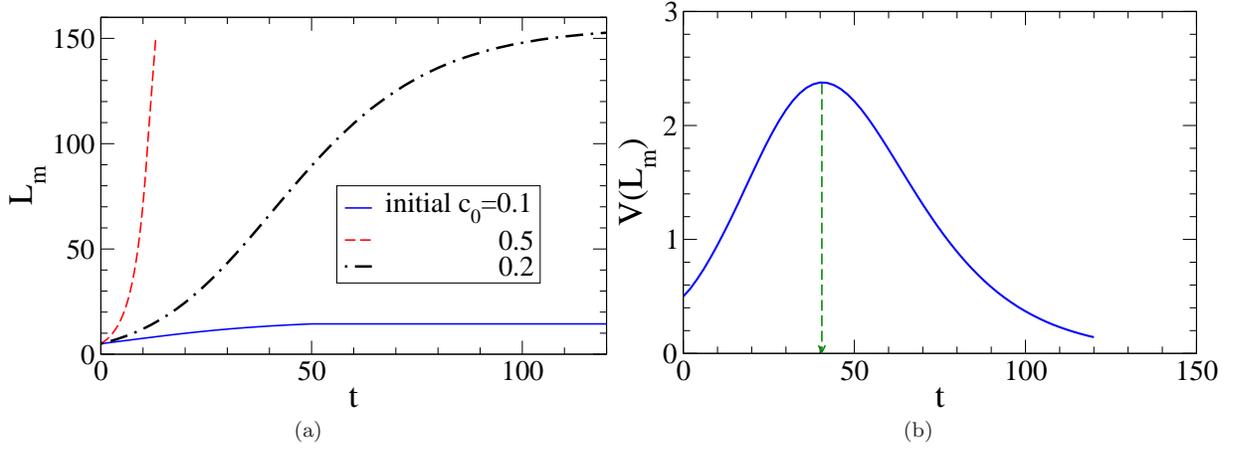

      \subfigure[]{\includegraphics*[width=\figws]{Lmmu1nu_5p_9.eps}} 
       \subfigure[]{\includegraphics*[width=\figws]{Vvstmu1nu_5p_9.eps}}
      \caption{Different tissue growth modes in the bistable region. (a) The tissue size $L_m(t)$ for the system in Fig. \ref{growth3} under three different initial values of $c_0$.     Time and $L_m$ are in  units of $1/d$ and $\sqrt{D/d}$ respectively. (b) The growth speed of the leading edge of the tissue as a function time in (a) for the initial $c_0=0.2$, showing a peak at a time $\tau_{sw}$ (marked by vertical dashed line) the accelerated growth is switched to retarded growth.}\label{Lmt}
      \end{figure}

  Fig.  \ref{Lmt}a shows the tissue size growth dynamics  in the bi-stable regime for different values of initial $c_0$. For initial $c_0=0.1$, the $L_m$ increases and saturates to the final tissue size at which the stem cell will extinct and the tissue growth then stopped. On the other hand, for initial $c_0=0.5$, $L_m$ shows a rapid exponential increase. More interestingly, for  
 initial $c_0=0.2$, $L_m$ displays a pronounced early stage of accelerated growth and later on switched to a retarded growth before it eventually approaches to the final-state, which was observed in a broad class of biological growths. Such S-shape growth is characterized by the existence of an inflexion point in $L_m(t)$ or equivalently there is a maximum of the instantaneous tissue growth speed as shown in the numerical results in Fig. \ref{Lmt}b.
 Furthermore, it should be noted that ultimate fate of the  tissue growth depends only on the initial value of $c_0$, but is independent of the initial tissue size $L_m(0)$ since the dynamics is governed by the phase flow as depicted in Fig. \ref{Ac0}c.

\subsection{Tissue growth time scales and final-state tissue size}
The growth modes of the tissue  to the final-state is governed by the presence of the stable trivial fixed point and the characteristic dynamics is determined by the approach rate to this stable fixed point which is predicted by (\ref{taus}). 
The saturation time-scale of the final-state growth of the tissue is then given by 
\begin{equation}
{\tau_s }^{-1} = \nu[1-\tfrac{2p}{1+{\tilde \mu}^m}],\label{tauss}
\end{equation}
which can be checked against the values obtained from the fitting of the numerical solutions.
 Fig. \ref{Lmvst}a displays the saturation dynamics of the tissue size to the final-state. The tissue size is well-fitted with the functional form
$L_m(t)=a_0+ a_1 e^{-\frac{t}{\tau_s}}$ (solid curves) from which  the predicted time scale $\tau_s$ can be extracted ($a_0$ and $a_1$ are fitting parameters also). The extracted final-state growth time scales are obtained as a function of  $p$ for two different values of ${\tilde \nu}$, and the results are shown in Fig. \ref{tauvsp}a. $\tau_s$ increases slowly with $p$ and is inversely proportional to ${\tilde \nu}$. The theoretical predictions Eq. ({\ref{tauss}) (curves) are also displayed showing very good agreement.

For the case of blow-up growth, our theory indicates that the growth dynamics is governed by  (\ref{LmODE}) giving rise to exponential growth. This is verified from the numerical results of $L_m(t)$ in  Fig. \ref{Lmvst}b from which the growth time scale $\tau$ can be extracted by fitting of the tissue dynamics in the long-time data.  The extracted exponential growth time scales are obtained as a function  of $p$ for two different values of ${\tilde \nu}$ are displayed in Fig. \ref{tauvsp}b, showing very good agreement as well.  $\tau$ decreases with $p$ and is inversely proportional to ${\tilde \nu}$.
\begin{figure}[htbp]
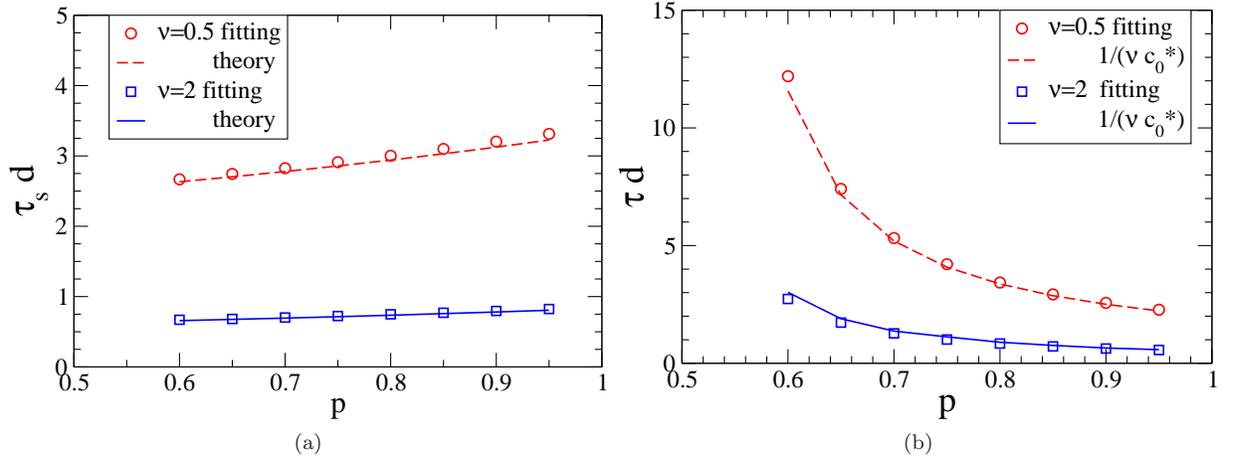

\centering
\subfigure[]{\includegraphics*[width=\figws]{tausvsp.eps}}
\subfigure[]{\includegraphics*[width=\figws]{tauvsp.eps}}
      \caption{Characteristic time-scales (in unit of $1/d$) for the  controlled growth and  blow-up growth, for $m=2$. Symbols are time-scales obtained from fitting of the tissue-size from the numerical solutions. Curves are the corresponding theoretical predictions. (a)$\mu=2$ for the case of controlled growth to the final-state. Theoretical predictions are from (\ref{tauss}). (b)$\mu=0.2$ for the case of blow-up growth. Theoretical predictions are from (\ref{tau}).}
\label{tauvsp}
\end{figure}

 As shown in previous section that it is possible for the approach to the final-state via an early acceleration and then follow by retarded growth. The condition for such a scenario to occur is given by (\ref{c0sw}) for the initial SC cell concentration to exceed some threshold value. For the Hill function regulation (\ref{PA}), the above  condition reads
 \begin{equation}
  c_0(0)> c_0^{sw}\equiv \frac{{\tilde \mu}-(2p-1)^{\tfrac{1}{m}}}{{\tilde \mu}+{\tilde \nu}(2p-1)^{\tfrac{1}{m}}}.\label{c0sw2}
  \end{equation} 
  Fig. \ref{LmtS}a shows the numerical results for the tissue size growth dynamics for such cases. The instantaneous leading edge speed displays a maximum at $t=\tau_{sw}$ signifying the switch from accelerated growth to retarded growth (see the veritcal dashed lines in Fig. \ref{LmtS}b). The dynamics of the SC concentration is also shown and the corresponding value at $t=\tau_{sw}$ is marked by a horizontal dot-dashed line at a value $ c_0^{sw}\simeq 0.05$, which is in reasonable agreement from the theoretical value of 0.047 from (\ref{c0sw2}).
\begin{figure}[htbp]
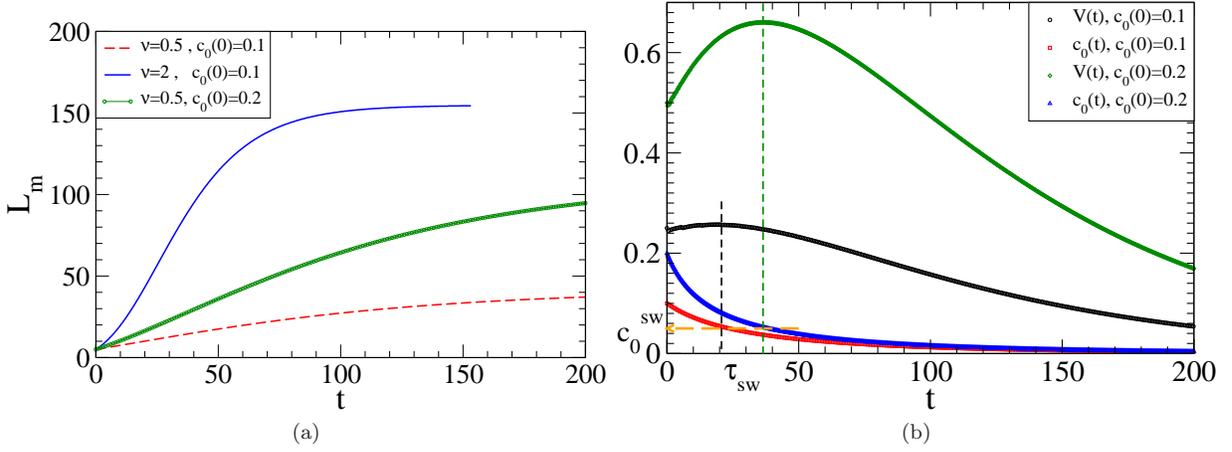

\centering
\subfigure[]{\includegraphics*[width=\figws]{Lmvstmu_48nu_5nu2p_6b.eps}}
\subfigure[]{\includegraphics*[width=\figws]{Vvstmu_48nu_5p_6b.eps}}
      \caption{Numerical results on the tissue size dynamics of the lineage model showing the S-shape growth curve, with $m=2$ for (a) Leading edge tissue size as a function of $t$ for  ${\tilde \mu}=0.48$, $p=0.6$ and  ${\tilde \nu}=0.5$ and 2. (b) The instantaneous growth speed of the leading edge of the tissue in (a) for  ${\tilde \nu}=0.5$ and initial SC cell density $c_0(0)=0.1$ and 0.2. The maximum of the speed is marked by  a vertical dashed line, which occurs at $t=\tau_{sw}$ with the corresponding SC cell density decay to $c_0^{sw}\simeq 0.05$ (marked by dashed horizontal arrow), which agree reasonably well with the theoretical prediction (from (\ref{c0sw2})) of  0.047 and is independent of the values of ${\tilde \nu}$.  Time and $L_m$ are in  units of $1/d$ and $\sqrt{D/d}$ respectively.}
\label{LmtS}
\end{figure}

 The ultimate tissue size can be achieved for the final-state growth is of practical interest.  From the numerical solutions, the tissue size saturates at long times and the ultimate size, $L_m(t\to\infty)$, can be measured. The time evolution of the tissue size is displayed in  Fig. \ref{Lmbigtfig}a, showing the saturation approach to the ultimate tissue size. The result indicates that $L_m(\infty)$ is not sensitive to the value of ${\tilde \nu}$. The theoretical approximation of $L_m(t)$ for the final-state dynamics given by (\ref{LmtBessel}) is also plotted, showing reasonable agreement. The ultimate tissue size of the final-state as a function of ${\tilde \mu}$ is shown in  Fig. \ref{Lmbigtfig}b, showing a decrease in ultimate size for increasing ${\tilde \mu}$. The theoretical estimations for $L_m(\infty)$ from (\ref{Lmbigt}) also show good agreement with the numerical results.
\begin{figure}[h]
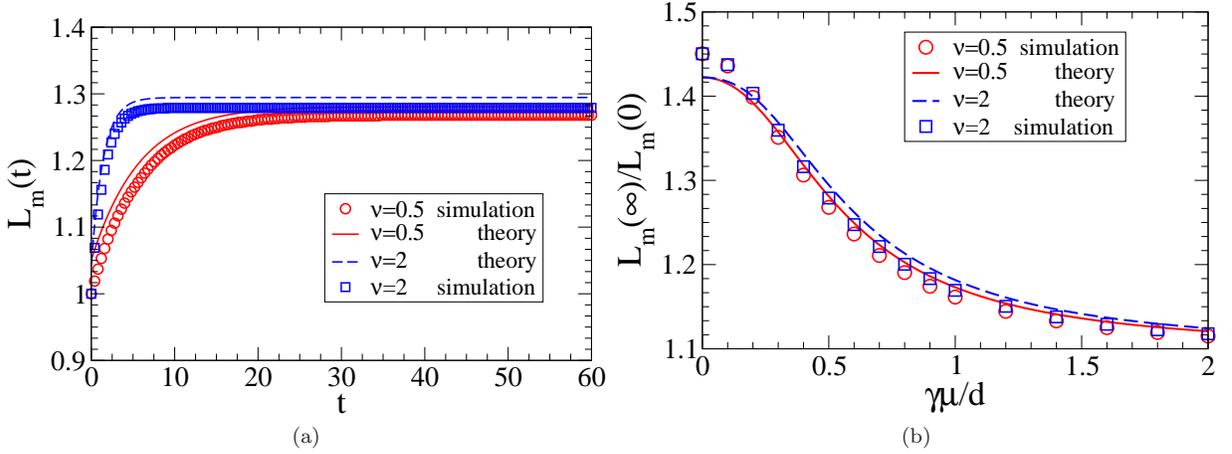

\subfigure[]{\includegraphics*[width=\figws]{LmtBesselp_4mu_5.eps}}
 \subfigure[] {\includegraphics*[width=\figws]{Lmbigtvsmup_4.eps}}
\caption{(a) Tissue size time evolution to the final-state from numerical solutions for ${\tilde \nu}=0.5$ and 2 (symbols) with  $c_0(0)=0.1$ and $p=0.4$. The theoretical approximations from (\ref{LmtBessel})  (with $f=0.5$) for ${\tilde \nu}=0.5$ (solid curve)  ${\tilde \nu}=2$ (dashed curve)are also shown.   Time and $L_m$ are in  units of $1/d$ and $\sqrt{D/d}$ respectively. (b) Normalized $L_m(\infty )$ as a function of ${\tilde \mu}$  measured from the numerical solutions at long times. The theoretical results from (\ref{Lmbigt}) are also shown. }
\label{Lmbigtfig}\end{figure}

The numerical solutions can provide detail quantitative results such as the detail concentration profiles of the leading edge of the growth, which is not easily obtainable analytically. In addition, the basin of attraction in the phase space (which can only be obtained numerically) can provide valuable information in the evolution dynamics of the growth and the sensitivity of external influences to alter the fate of the growth.

\section{Some Possible Applications}
Although the model considered in previous section is rather simple, it can be applied to various experimental or clinical scenarios to provide insights for practical purposes. A few cases are considered below. 
 
  \subsection{Growth mode switching with regulatory pulse control}
 The two major growth modes in our model are blow-up growth and final-state growth, whose properties are governed respectively by the corresponding non-trivial and trivial stable fixed points. Moreover, in the bistable regime in which these two growth modes can coexist, one can externally perturb the system and drive one mode to the other and vice versa. This can be achieved by controlling the concentration of the regulatory molecules externally.  We demonstrate this in the bistable regime as that in  Fig. \ref{growth3}. Fig. \ref{switch} shows that different growth modes can be switched to one another in the bistable regime. In Fig. \ref{switch}a, the original final-state mode (dashed curve) is switched to the blow-up mode when a pulse control, which keeps the concentration $A$ to a low level for a fixed duration, is applied. Carefully examination of the evolution of the values of $c_0$ and $A$ after the pulse control indicates that the dynamics indeed flows to the non-trivial fixed point.
 Conversely, as shown in Fig. \ref{switch}b, the explosive size increase in an original blow-up growth can be suppressed
 by a similar pulse control that keeps $A$ to a high level to suppress the subsequent growth to a final-state. Quantitative knowledge on the characteristic time scales of the blow-up and final-state growth dynamics as described in Sec. IVB is essential in designing the pulse control duration and the timing to applied for a successful growth mode switching.
\begin{figure}
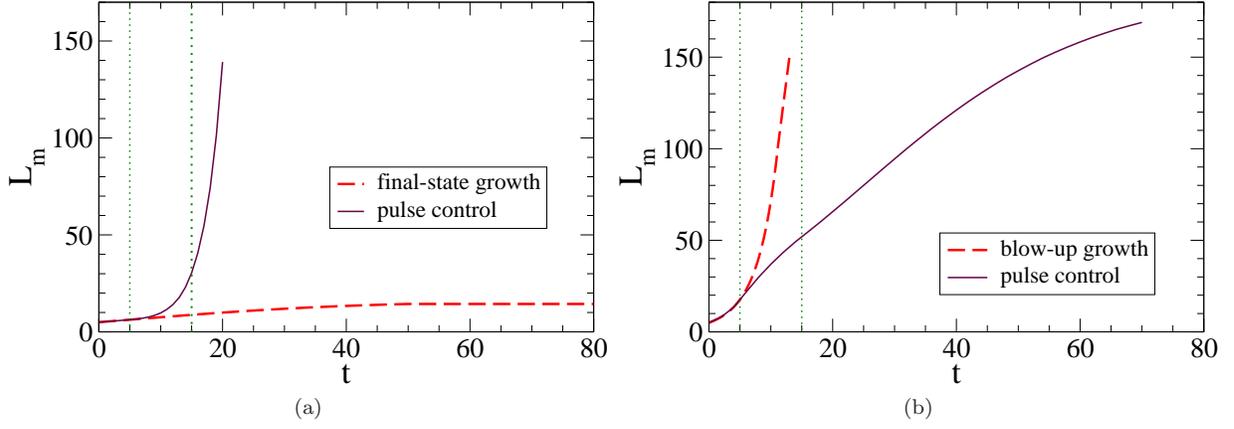

\begin{center}
 \subfigure[]{\includegraphics*[width=\figws]{Lmmu1nu_5p_9control2.eps}}
 \subfigure[]{\includegraphics*[width=\figws]{Lmmu1nu_5p_9contorl1.eps}}
    \end{center}  \caption{Switching of growth modes by pulse control of the regulatory molecular concentration. (a) Tissue size growth dynamics with the initial growth condition as in Fig. \ref{growth3}a. Pulse control of $A$ is applied for $5\leqslant t \leqslant 15$ (between the vertical dotted lines) by keeping $A=0.2$ in this period. The original final-state growth is boosted to the blow-up mode. (b) Tissue size growth dynamics with the initial growth condition as in Fig. \ref{growth3}c. Pulse control of $A$ is applied for $5\leqslant t \leqslant 15$ by keeping $A=1$ in this period. The original blow-up growth is suppressed to the final-state mode.  Time and $L_m$ are in  units of $1/d$ and $\sqrt{D/d}$ respectively.}
 \label{switch}
 \end{figure}
 
  \subsection{Controlability and Engineered Linear growth}
  Understanding the growth dynamics of the system enable one to  design the desired tissue growth mode by controlling the concentration of the regulatory  molecule with external feedback, i.e.  adjusting $A$ by adding or depleting  regulatory molecules with real-time feedback according to the instantaneous stem cell population. Here we demonstrate such an external  feedback control design to achieve a linear growth mode. For the tissue to grow linear with time, one requires $\frac{d^2 L_m}{d^2t}=0$ and  (\ref{Lmacc}) tells us that this can be achieved by adjusting $A$ such that $2{\cal P}(A)=1$ at all times. For ${\cal P}(A)$ given by (\ref{PA}), to achieve a linear growth, one simply designs the feedback  to maintain the value of $A={\tilde \mu}_0(p)/\gamma$ at all times. One first needs to know the intrinsic concentration of regulatory molecules, $A_{intrin}$ secreted by the cell. $A_{instrin}$ can be estimated by assuming step function profiles for both $c_0$ and $A_{instrin}$. Using (\ref{PDE3}) for USS, one has $A_{intrin}\simeq \frac{\mu(1-c_0)}{d(1+{\tilde \nu}c_0)}$, thus one needs to increase the concentration of the regulatory molecule  externally  by an  amount $ A_{ext}=[{\tilde \mu}_0(p)- \frac{{\tilde \mu}(1-c_0)}{(1+{\tilde \nu}c_0)}]/\gamma$.
  
  The above feedback control is implemented in numerical simulations and the results are displayed in Fig. \ref{growthlinear} showing the success of achieving the linear growing tissue size. It should be noted that under the linear control, the growing speed is proportional to ${\tilde \nu} L_m(0)$, but is independent of ${\tilde \mu}$ and $p$.
  \begin{figure}
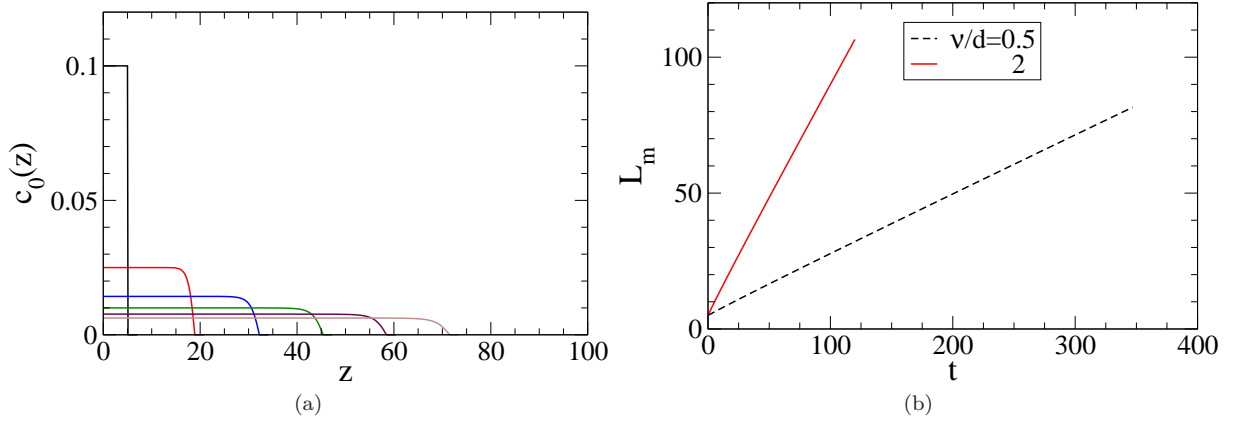

\begin{center}
\subfigure[]{\includegraphics*[width=\figws]{c0zmu_2nu_5p_6linear.eps}}
 \subfigure[]{\includegraphics*[width=\figws]{Lmtmu_2p_6lineaer.eps}}
    \end{center}  \caption{Tissue designed to grow linearly by feedback control of the concentration of the regulatory molecules.   $m=2$. (a) Time evolution of the $c_0(z)$ profiles for  ${\tilde \nu} = 0.5$, each curve is separated by  time intervals of 60 units.   Time and space are in  units of $1/d$ and $\sqrt{D/d}$ respectively.
    (b) Tissue size as a function of time showing the linear growth behavior for ${\tilde \nu} = 0.5$ and 2.   Time and $L_m$ are in  units of $1/d$ and $\sqrt{D/d}$ respectively.}
 \label{growthlinear}
 \end{figure}
 
 \subsection{Catch-up growth}
 Catch-up growth is often observed in child development\cite{lucas}. After a period of growth retardation caused by severe illness, subsequent acceleration of the growth rate can occur which involves rapid increase in weight or length in infants until the normal individual growth pattern is resumed. This phenomenon has been studied for hundred of years, but the mechanism of growth transition is not very clear, and strategy in applying external growth stimulations(by hormones or growth factors) to achieve effective catch-up will be desirable. Using our theory, we can model  the catch-up growth phenomenon and give some insight for the strategic implementation. We model  the catch-up growth  by a transient duration of increase in the progenitor cell cycle speed, i.e. the  proliferation rate parameter ${\tilde \nu}$ in our model, while all other parameters remain unchanged. The analytic phase diagram in Fig. \ref{phase} can provide valuable insights to determine whether a catch-up growth is possible by merely increasing ${\tilde \nu}$.  For catch-up growth, one looks for a final-state upon which an increase in ${\tilde \nu}$ can change it to the blow-up state. Careful examination of the two phase diagrams in Fig. \ref{phase} reveals that to switch from a final-state (region II) to a pure blow-up state (region I) is impossible because the phase boundary separating these two states is ${\tilde \mu}_0(p)=(2p-1)^{\frac{1}{m}}$ (the dotted curve) which is independent of ${\tilde \nu}$. However, the phase boundary between the final-state (region II) and the bistable coexistence state (region III) does depend on ${\tilde \nu}$.
For instance, if one  chooses the original final-state  with   $p=0.8$,  ${\tilde \mu}=1$ and ${\tilde \nu}=0.5$  (see Fig. \ref{phase}a). Then with an increase of ${\tilde \nu}$ to 2 during the catch-up period, the system is switched to the bistable growth state (see Fig. \ref{phase}b) and hence allowing the possibility of a blow-up growth for catching up. In addition, the stem cell concentration $c_0$ cannot be too small so that blow-up growth will occur in the bistable state. Fig.  \ref{catch}a demonstrates the success of a catch-up growth with  the original final-state of  $p=0.8$,  ${\tilde \mu}=1$, $c_0=0.3$ and ${\tilde \nu}=0.5$; followed by a short  catch-up duration of 2 time units during which ${\tilde \nu}$ is switched to 2. As shown in Fig.  \ref{catch}a, if the catch-up period occurs at an early stage ($5\leqslant t \leqslant 7$), then the catching up is rather successful with a final tissue  more than twice as the of the original growth. On the other hand, if the catch-up occurs later  ($10\leqslant t \leqslant 12$), the effect of catch-up growth is much less pronounced.
 
  \subsection{Timing in growth boosting} Here we assume the boosted growth is initiated by some upstream regulation or external stimulations that suppress the secretion rate of the regulatory molecules from the  TD cells, i.e. a decrease in $\mu$, for a period of time so that the growth can be faster to catch up.  To model such a catch-up period, we impose a pulse control of a small value of $\mu$  for some period of time in the original final-state growth.  Fig. \ref{catch} shows  the tissue size dynamics for an initial final-state growth (region II of the phase diagram in Fig. \ref{phase}a), the boosted growth is applied at  an early and a later times. During the boosted period, the value of ${\tilde \mu}$ is kept at a low value such that the system is pushed to region I of the phase diagram in Fig. \ref{phase}a for blow-up growth. In practice, to lower the value of ${\tilde \mu}$ can be achieved by decreasing the production rate $\mu$ or  by increasing the decay rate $d$ of $A$, or by decreasing the regulation strength $\gamma$. Furthermore, Fig. \ref{catch} also suggests that effective boosted growth can be achieved if it is applied at an early stage (solid curve), otherwise if the system has already grown near to its final-state, the same boosting duration has little effects on the final size of the tissue (dot-dashed curve).  In other word,  the timing, duration and strength of the boosted pulse are all essential in determining the ultimate mature size after the boosted period.
\begin{figure}
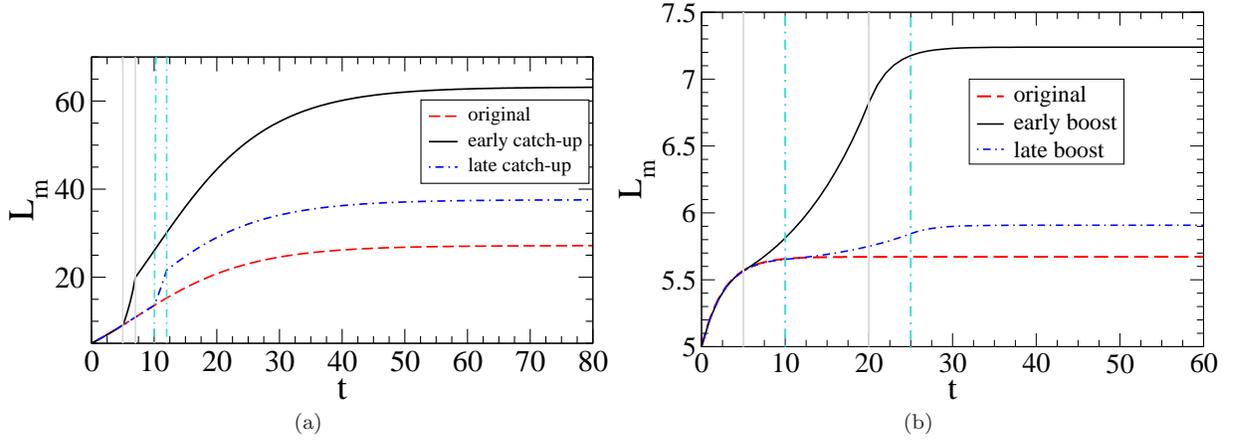

\begin{center}
\subfigure[] {\includegraphics*[width=\figws]{Lmcatch.eps}}
\subfigure[] {\includegraphics*[width=\figws]{Lmmu2nu_5p_6boost.eps}}
  \end{center}  \caption{(a) Catch-up growth modeled by a short duration of increase in the progenitor cell proliferation rate parameter ${\tilde \nu}$.
  The original final-state grows with   $p=0.8$,  ${\tilde \mu}=1$, $c_0=0.3$ and ${\tilde \nu}=0.5$ (dashed curve). A catch-up growth occurs at an early stage (solid curve) in   $5\leqslant t \leqslant 7$ (between two vertical solid  lines) during which ${\tilde \nu}=2$. Another similar catch-up growth occurs at a later stage (dot-dashed curve) in   $10\leqslant t \leqslant 12$ (between two vertical dot-dashed  lines) during which ${\tilde \nu}=2$.  (b) Boosted growth modeled by a pulse control of the secretion rate parameter of the regulatory molecules, ${\tilde \mu}$. The original growth of the tissue is
  with parameters $m=2$, ${\tilde \nu} = 0.5$, ${\tilde \mu}=2$ and $p=0.6$.  The tissue size growth curves are shown for the original (dashed curve), early boosted (solid curve), and late boosted (dot-dashed curve) growths. An early  pulse control of ${\tilde \mu}$ is applied for  $5\leqslant t \leqslant 15$ with the change ${\tilde \mu=0.1}$ in this period.
  Another similar boosted pulse control is also applied but at a later time in  $10\leqslant t \leqslant 20$. The vertical lines show the corresponding boosted periods during which ${\tilde \mu}$ is changed.   Time and $L_m$ are in  units of $1/d$ and $\sqrt{D/d}$ respectively.}
 \label{catch}
 \end{figure}

\section{Conclusion and Outlook}
A  two-stage lineage cell model with spatially diffusive negative feedback signalling molecules,   focusing on the  tissue growth dynamics, is investigated analytically and numerically. By deriving the fixed points for the uniform steady states and carrying out linear stability analysis,  the phase diagrams are obtained analytically for arbitrary parameters of the model for Hill function form of the negative feedback.
Different growth modes, including saturation dynamics to a final-state of finite tissue size and  blow-up growth in which the tissue size become exponentially diverging, are obtained in the model. The rich growth dynamics is summarized in terms of the tissue size $L_m$ as follows: the final-state growth mode occurs in region II of the phase diagram, characterized with retarded growth, $\frac{d^2 L_m}{d^2t} <0$, at late times. And blow-up growth mode occurs in region I of the phase diagram with accelerated growth $\frac{d^2 L_m}{d^2t} >0$. And in the bi-stable regime III of the phase diagram, depending on the initial $c_0$ concentration, the ultimate dynamics can approach a final-state tissue size or grow exponentially. On top of the above,
 it is possible to have a typical biological growth mode of an accelerated early growth and a retarded growth at late stage  (a a S-shape tissue growth curve), characterised by the presence of a switching time $\tau_{sw}$ at which  $\frac{d^2 L_m}{d^2t}=0$ (inflexion point) with a maximal instantaneous growth speed (see Fig. \ref{Lmt}c), if the system lies in region above the phase boundary  ${\tilde \mu} > {\tilde \mu}_0(p)$  and the initial stem cell density is sufficiently high (as given by (\ref{c0sw})).
 Furthermore, the existence of a bistable regime for a wide range of parameters would provide a buffered regime for  the controlled growth to finite tissue size against possible stochastic fluctuations that could otherwise lead to uncontrolled blow-up growth. 
 It is worth to compare our model with some similar model such as in  \cite{Lander2016}in which both positive and negative feedback are consider in a two-lineage model but without the explicit dynamics of a regulatory molecule ($A$ in our model). For the special case of only negative feedback with Hill coefficient $m=1$ in \cite{Lander2016}, bistable regime will not exist, in agreement with the prediction in our model. In addition, our model  also gives the explicit result of $p_c({\tilde \nu})$ (Eq. (\ref{pstarnu})) quantifying the termination of the bistable region for $p<p_c({\tilde \nu})$, and also predict the final-state tissue size $L_m(\infty)$ in (\ref{Lmbigt}).

 Several illustrative studies are also carried out to demonstrate the possibility of applying our model to the growth control strategy, including the controlled switching between final-state growth and blow-up growth,  the appropriate timing for effective boosted growth and catch-up growth, and the design to achieve a target  growth dynamics such as the engineered linear growth. It is demonstrated that the knowledge of the  analytic phase diagrams such as those in Fig. \ref{phase} is very valuable for the success of implementing the growth control.
 
It is worth to note that in other studies of feedback-driven morphogenesis,  both positive and negative feedbacks are required to achieve the bi-stability. But in our model, mere negative feedback regulation on the proliferation of stem cells can realize the bi-modal growth.
In support of our finding that just negative feedback is sufficient for achieving growth bimodality, we note the following  phenomenon was reported in experiments. In the mouse olfactory epithelium(OE), a reduction in the strength of FGF signaling (due to loss of Fgf8\cite{kawauchi2005}), can lead to not just  a smaller OE, but even to a complete absence of tissues(agenesis). While agenesis due to loss of Foxg1 can be rescued by inactivation of Gdf11\cite{kawauchi2009}, interestingly, inactivation of even a single allele of Gdf11 in Foxg1 mutant animals can restore the OE to full thickness. Such experimental findings were consistent with idea that growth modes can be switched sensitively by negative feedback. 
In  our  model, the properties of the negative feedback can be varied by changing the parameters $\gamma$, $\mu$, or $d$, i.e. the regulation strength, feedback production and death are all closely related to effectiveness of the feedback and takes parts in the switching of the growth modes. Our findings can give  further inspirations on biological experiments that there may be more diverse channels for the  control of the growth of cell lineages and tissue sizes.
 
In the theoretical analysis of the stability of the spatially homogeneous solution, we focus on the interior region of the tissue which is far away from the leading edge. For a steadily growing tissue, the length scale of the leading edge with significant concentration gradient is small compare to the bulk tissue, and hence we only focus on the stability of the bulk tissue in the theoretical analysis. In reality or in the numerical solution, the concentration gradient near the leading edge will lead to deviation from the uniform concentration profile, as seen in the numerical plots. Also in Sec. IV, the spatial profiles are assumed to be step-functions to allow for the analytic results on the tissue size dynamics. Since the tissue size is mainly dominated by the growth in the bulk (which depends on the bulk concentration), the assumption of a step-function profile (neglecting the shape profile of the leading edge) is reasonable. Such assumptions in the theoretical analysis can be justified from the fact that the results of the tissue growth dynamics measured from the numerical solutions agree well with the analytic predictions. On the other hand, one needs to resort to the numerical solution for the detail concentration profile of the leading edge.  
 
 In our model, the stem cell might become extinct in the parameter regime of the final-state growth. On the other hand, due to changes in environmental stimuli or changes in internal/external conditions, the  biological system might need to adapt and to re-start to grow again, such as in the case of fast tissue regeneration. Then the system needs to be regulated by other upstream pathways that would lead to the revival of the stem cell production, together with the regulations that cause the change or switch of the parameters in the current status. Such a possibility of adaptability to change the growth pattern can be extended in the present framework by including possible upstream regulatory pathways.
 
One spatial dimension is considered in this work in the theoretical model (\ref{PDEG1}), which  can be extended to higher dimensions by $\partial_z (V c_0)  \to \nabla\cdot ({\vec V}c_0)$ and $\partial_z^2 \to \nabla^2$ ,  with a growth velocity vector ${\vec V}$.
For higher spatial dimensions, the same USS solutions hold as in the one-dimensional case, i.e. the same trivial and non-trivial fixed points will govern the growth fate of the system, and hence one expects the conclusion in the present work is expected to hold qualitatively also.
For the simple case that ${\vec V}$ is along the (outward) normal direction of the tissue boundary, similar numerical schemes (as outline in Appendix  II) can be applied as in the one-dimensional case with the extra complication of updating a moving domain boundary, and the resulting dynamics is qualitatively similar. 
In general, the growth velocity (direction and magnitude) is determined by the tissue mechanics and constitutive equations of the cells niches and tissues. For instance, ${\vec V}$ can be assumed to be the passive velocity governed by some generalized Darcy's law as a result of the pressure induced by cell proliferations as well as determined by the fluxes due to the interactions among the cells. The resulting growth of the tissue, cell concentration and pressure field can also be solved numerically. However, due to the complex interactions and the interplay of soft-tissue mechanics, new spatial instabilities might arise that could lead to spatial patterns which is an interesting problem to be explored in details.

 The present work focused on a simple two-stage lineage cell model, it can also be extended to include lineage of multiple stages\cite{Lander2009} or branched lineages\cite{Lander2015} with  cross-regulations across different lineages. The interplay between self-proliferation, differentiation and de-differentiation\cite{jychang,mxwang}, cell-cell interactions\cite{wanglai2012,wanglai2018} can be incorporated to investigate the effects on the growth dynamics. 
 With the present theoretical basis, more sophisticated clinical situations can be modeled with appropriate extension of the present model. For example  in cell transplantation, the growth dynamics for a  transplanted new growing bud in a mature tissue is the focus of regenerative medicine. The transplantation experiments  in mouse muscle showed that the myofiber associated satellite cells are allowed to re-populate injured muscle\cite{hall}. In such experiments the transplanted cells, which are FGF2-treated prior to transplantation, trigger an abnormally high rate of myoblast proliferation and differentiation, which can be sustained without further intervention for years. Our model can be extended to include two types of stem cells with different parameters corresponding to the (original) final-state and  blow-up state (for the transplanted cells). The above transplantation growth dynamics  can be modelled by coupled partial differential equations of the two stem cell concentrations and the feedback molecular concentrations.

{\bf Acknowledgement }
     MXW thanks the funding support from National Natural Science Foundation of China (Grant No.  11204132). AL acknowledges the support of National Institutes of Health (Grant No. R01-NS095355).      PYL thanks Ministry of Science and Technology of Taiwan under grant  110-2112-M-008-026-MY3 and National Center for Theoretical Sciences of Taiwan for support.
\newpage

\section*{Appendix I: Uniform steady-states and their Stability }
\subsection{General ${\cal P}(A)$ and  Linear stability analysis} First we consider general negative feedback regulatory function ${\cal P}(A)$ which is assumed to be a monotonic decreasing function.
The  uniform steady-state (USS) is given by the equations for the fixed points 
\begin{eqnarray}
 c_0^2&=&(2{\cal P}(A)-1)c_0\\
 \nu c_0 A&=& \mu(1-c_0)- A d.
\end{eqnarray}
One can see easily that the trivial fixed point $(c_0,A)=(0,\mu/d)$ always exists, and  the non-trivial fixed point(s) $(c_0*\neq 0, A^*)$ may exist which is  given by the root of the following equation for $A^*$:
\begin{equation}
2 (\mu+\nu A){\cal P}(A)=2\mu+(\nu-d)A,\label{Astar}
\end{equation}
and $c_0^*$ is given by
\begin{eqnarray}
c_0^*&=&2{\cal P}(A^*)-1\\
&=& \frac{\mu-A^*d}{\mu+\nu A^*}\label{C0star}.
\end{eqnarray}
It follows that for physical non-trivial solution of $1\geqslant c_0^*>0$, one has ${\cal P}(A^*)>\tfrac{1}{2}$, and using  (\ref{Astar}), it is also equivalent to
\begin{equation}
0\leqslant A^*<\frac{\mu}{d}.\label{Astar2}
\end{equation}

Notice that although in general the non-trivial fixed point $c_0^*\neq 0$, it can be seen from (\ref{Astar}) and (\ref{C0star}) that $c_0^*= 0$ at the specific value of  (${\cal P}^{-1}$ always exists since ${\cal P}(A)$ is a monotonic decreasing function)
\begin{equation}
\mu=\mu_0\equiv {\cal P}^{-1}(\tfrac{1}{2})d.\label{mu0}
\end{equation}

The stability of the fixed point can be analyzed by considering small deviations from the USS $(c_0^{(u)}, A^{(u)})$ with
$c_0\simeq c_0^{(u)} +\delta c_0$ and $A\simeq A^{(u)}+\delta A$. Then (\ref{PDE2}) is linearized to give
\begin{equation}
\frac{ \partial}{\partial t}\colvec{ \delta c_0} {\delta A} =- \begin{pmatrix}\nu c_0^{(u)}(1+z\partial_z)-\nu(2{\cal P}(A^{(u)})-1) & -2\nu c_0^{(u)} {\cal P}'(A^{(u)})\\\mu+\nu A^{(u)}& \nu c_0^{(u)}(1+z\partial_z)+d-D\partial^2_z\end{pmatrix}\colvec {\delta c_0} {\delta A}.\label{delC0}
\end{equation}

\underline{For the trivial fixed point of $(0,\frac{\mu}{d})$}, and deviation with spatial wavenumber $q$, i.e. ${\delta A}\sim e^{i q z}$, the eigenvalues of the Jacobian matrix in (\ref{delC0}) can be calculated to  be $-(d+Dq^2)$ and $\nu(2{\cal P}(\frac{\mu}{d})-1)$. Hence the trivial fixed point will be stable for all wavelength if ${\cal P}(\frac{\mu}{d})<\tfrac{1}{2}$, and becomes unstable for ${\cal P}(\frac{\mu}{d})>\tfrac{1}{2}$. Since ${\cal P}(A)$ is a monotonic decreasing function, this implies that the trivial state is unstable for small values of $\frac{\mu}{d}$, but becomes stable for sufficiently large $\frac{\mu}{d}$. In addition, a stable trivial fixed point of $c_0=0$ corresponds to the controlled growth of the tissue whose size approach a final-state saturated value. The saturation rate of final-state growth tissue is given by the rate of $c_0$ approaching the trivial USS fixed point, and can be estimated from the  corresponding Jacobian matrix
whose eigenvalues and eigenvectors can be calculated to give
\begin{eqnarray}\lambda&=&-d , \quad (0,1)\\
\lambda&=& -{ \nu}[1-2{\cal P}(\tfrac{\mu}{d})],\quad (\frac{{\tilde \nu}[1-2{\cal P}(\tfrac{\mu}{d})]-1}{ \tfrac{\mu}{d}(1+{\tilde \nu})},1).
\end{eqnarray}
 Since the eigenvector corresponding to $\lambda=-d$ has no component along the $c_0$ axis, thus the asymptotic dynamics of $c_0$ relaxing to the stable $c_0=0$ final-state is governed by the other eigenvalue. Hence the corresponding saturation time-scale ($\tau_s$) for the final-state growth is then given by 
\begin{equation}
\tau_s^{-1}=\nu[1-2{\cal P}(\tfrac{\mu}{d})].\label{tausApp}
\end{equation}

\underline {For the non-trivial fixed point of $(c_0^*\neq 0,A^*)$}, with  $\delta c_0$ and  ${\delta A}\sim e^{i q z}$, the Jacobian matrix from (\ref{delC0}) is
\begin{equation}
{\bf J}^*=\begin{pmatrix}-\nu c_0^*(1+i qz) & 2\nu c_0^* {\cal P}'(A^*)\\
-\mu-\nu A^*& -\nu c_0^*(1+i qz)-d-Dq^2\end{pmatrix}.\label{J}
\end{equation}
Careful analysis reveals that the real part of the eigenvalue of ${\bf J}^*$ is independent of the imaginary term $iqz$, and hence the stability of the non-trivial USS is determined by  ${\bf J}^*|_{iqz=0}$, whose trace and determinant are given by 
\begin{eqnarray}
\hbox{Tr}&=&-2\nu c_0^*-d-Dq^2\label{Tr}\\
\det&=&\nu c_0^*[\nu c_0^*+d+Dq^2+2(\mu+\nu A^*) {\cal P}'(A^*)].\label{detq}
\end{eqnarray}
 Since Tr$<0$ and hence the non-trivial USS is stable (unstable) if det$>0$ ($<0$).
From (\ref{mu0}) and (\ref{detq}), it follows that at $\mu=\mu_0$ is always a stability boundary since $c_0^*$ (and hence the determinant also) changes sign on it.

\subsection{Fixed Points of the Uniform steady-states and stability analysis for ${\cal P}(A)=\frac{p}{1+(\gamma A)^m}$}
Hereafter, we shall consider the case with  ${\cal P}(A)=\frac{p}{1+(\gamma A)^m}$, where $m>0$ and is usually taken to be positive integer as a Hill coefficient. First for $p < \tfrac{1}{2}$, the trivial USS of $(0,\tfrac{\mu}{d})$ is the only fixed and there is no non-trivial fixed point of $c_0^* >0$. Notice that for non-trivial uniform state with $c_0^*>0$ (hence $1\geqslant {\cal P}(A^*)>\tfrac{1}{2}$), $1\geqslant p>\tfrac{1}{2}$.
 (\ref{Astar}) and (\ref{C0star}) reveal that
$c_0^*$ and $\gamma A^*$ are determined only by the following four positive parameters: ${\tilde \mu}\equiv\frac{\gamma\mu}{d}$, ${\tilde \nu}\equiv\frac{\nu}{d}$, $m>0$ and $p> \tfrac{1}{2}$.  In fact,  it is also clear that the behavior of the dynamical system (\ref{PDEG1}) (by choosing the time and space in  units of $1/d$ and $\sqrt{D/d}$ respectively) is also governed solely by these 4  dimensionless parameters. In particular, varying the parameter ${\tilde \mu}$ leads to interesting bifurcation behavior as will be shown in Fig. \ref{bifurcation}.

\subsubsection{Trivial uniform steady-state and its stability}
The trivial fixed point $(c_0,A)=(0,\mu/d)$ always exists and is independent of the form of ${\cal P}(A)$, but the Jacobian matrix depends on ${\cal P}(A)$ and in this case is simply
\begin{equation}
{\bf J}_0=d \begin{pmatrix}{\tilde \nu}\left( \frac{2p}{1+{\tilde \mu}^m}-1\right) & 0\\
-\gamma{\tilde \mu}(1+{\tilde \nu})& -1\end{pmatrix}.\label{J0}
\end{equation}
Hence  for $p < \tfrac{1}{2}$, the trivial fixed point is always stable. And for  $\tfrac{1}{2}\leqslant p \leqslant 1$, the trivial USS will be stable (unstable) if  $1+{\tilde \mu}^m > 2p$ ($<2p$), or the stability boundary of the trivial USS is
\begin{equation}
 {\tilde \mu}={\tilde \mu}_0(p)\equiv (2p-1)^{1\over m}\label{mu0p2}.
\end{equation}

\subsubsection{Number of non-trivial uniform steady-states}
For  ${\cal P}(A)=\frac{p}{1+(\gamma A)^m}$, it is convenient to define $X\equiv \gamma A^*$, and   from (\ref{Astar}) the non-trivial fixed point $\gamma A^*$  is given by the root of
\begin{equation}
F(X)\equiv (1-{\tilde \nu}) X^{m+1}-{2{\tilde \mu}}X^m+(1-{\tilde \nu}+{2p{\tilde \nu}})X -{2{\tilde \mu}(1-p)}=0.\label{FXFX}
\end{equation}
For positive integer values of $m$, $F(X)$ is a polynomial of degree $m+1$ (or degree $m$ if ${\tilde \nu}=1$), and we shall examine the possible number of positive roots below.  Direct calculations give
$F(0)=-2{\tilde \mu}(1-p) \leqslant 0$ and $F'(0)=1+(2p-1){\tilde \nu} > 1$ (since $p>\tfrac{1}{2}$). Furthermore, one has
\begin{eqnarray}
F'(X)&=& (1-{\tilde \nu})(m+1) X^{m}-{2m{\tilde \mu}}X^{m-1}+1-{\tilde \nu}+{2p{\tilde \nu}}\label{Fp}\\
F''(X)&=& mX^{m-2}[(m+1)(1-{\tilde \nu})X-2(m-1){\tilde \mu}].\label{Fpp}
\end{eqnarray}
There will be an inflexion point (i.e. $F''=0$) for $X$ at $X=\frac{2(m-1){{\tilde \mu}}}{(m+1)(1-{\tilde \nu})}$.
Hence there is a single inflexion point on the positive $X$-axis for ${\tilde \nu}<1$, while there is no inflexion point for ${\tilde \nu}\geqslant 1$. Therefore it follows that there can only be 1 or 3 non-trivial  positive root(s) for ${\tilde \nu}<1$ and none or 2 non-trivial  positive root(s) for ${\tilde \nu}\geqslant 1$.
Remarkably, the number of positive roots depends on the range of values of ${\tilde \mu}$ and can be calculated analytically as follows.  The critical value, ${\tilde \mu}_t$ at which the number of positive roots changes from 1 to 3 (for ${\tilde \nu} <1$) or 0 to 2 (for ${\tilde \nu} \geqslant 1$) can be obtained from the solution of 
$F(X_t)=0$ and $F'(X_t)=0$, where $X_t$ is the corresponding value of the root at ${\tilde \mu}_t$. Using (\ref{FX}) and (\ref{Fp}), with $Y\equiv X_t^m$ and  after some algebra, one can show that $Y$ satisfies a quadratic equation 
\begin{equation}
(1-{\tilde \nu})Y^2-[mp(1+{\tilde \nu})+(2-3p){\tilde \nu}+p-2]Y+(1-p)(1-{\tilde \nu} +2p {\tilde \nu})=0\label{Y2}
\end{equation}
whose solution $Y_{\pm}$ is given by
\begin{eqnarray}
2(1-{\tilde \nu})Y_{\pm}&=&[mp(1+{\tilde \nu})+(2-3p){\tilde \nu}+p-2]\nonumber
\\& &\pm \sqrt{[mp(1+{\tilde \nu})+(2-3p){\tilde \nu}+p-2]^2+(1-p)(1-{\tilde \nu}+{2p{\tilde \nu}})}\label{Ypm2}
\end{eqnarray}
with the corresponding ${\tilde \mu}_t$ given by
\begin{equation}
2{\tilde \mu}_t^{\pm}=\frac{m(1-{\tilde \nu})Y_{\pm}^{1+{1\over m}}}{(m-1)Y_{\pm}-1+p}.\label{mutpm2}
\end{equation}
Further analysis reveals that for the  case of ${\tilde \nu} <1$, there are 3 positive roots for ${\tilde \mu}_t^{+}<{\tilde \mu}<{\tilde \mu}_t^-$, and 1 positive root otherwise. ${\tilde \mu}_t^{+}$ and ${\tilde \mu}_t^{-}$ approach each other as $p$ decreases and there is a threshold $p_t({\tilde \nu})$ below which the 3-root regime vanishes. $p_t({\tilde \nu})$ can be calculated by setting the square root in (\ref{Ypm}) to zero to give
\begin{equation}
p_t({\tilde \nu})=\frac{4m(1-{\tilde \nu})}{(1+m^2)(1+{\tilde \nu})+2m(1-3{\tilde \nu})}.\label{pcnu}
\end{equation}

 For ${\tilde \nu} \geqslant 1$, there are 2 positive roots for ${\tilde \mu}<{\tilde \mu}_t^-$, and no positive root otherwise.
 The number of positive roots for the special case of $m=1$ can also be figured out directly.
The results for the  number of positive roots in $F(X)$ with the corresponding conditions on the values of ${\tilde \mu}$ are summarized in Table I.
\begin{table}[t]
\centering
\begin{tabular}{|c||c|c|}
\hline
 &  $m > 1$  & $m= 1$\\
\hline\hline
${\tilde \nu}<1$& 3 if  ${\tilde \mu}_t^{+}<{\tilde \mu}<{\tilde \mu}_t^-$, 1 otherwise & $1$ \\
\hline
${\tilde \nu}>1$& 2 if ${\tilde \mu}<{\tilde \mu}_t^-$, 0 otherwise & 2 if ${\tilde \mu}< {\tilde \mu}_t^-$, 0 otherwise\\\hline
 ${\tilde \nu}=1$&2 if ${\tilde \mu}<{\tilde \mu}_t^-$, 0 otherwise& 1 if ${\tilde \mu} < p$, 0 otherwise\\\hline
\end{tabular}
\caption{Table for the number of real and  positive roots of $F(X)$ in (\ref{FX}). ${\tilde \mu}_t^\pm$ are given by (\ref{Ypm}) and (\ref{mutpm})}\label{tabroots}
\end{table}
  It should be noted the non-trivial states given by  the positive roots in Table \ref{tabroots} need to comply with the physical requirement of $c_0^*>0$, namely $X< {\tilde \mu}$. As the value of  ${\tilde \mu}$ changes, the possibility of the emergence of new  fixed points in pairs (via saddle-node bifurcations) can lead to interesting transition for new  states, as will be explored below.

\subsubsection{Stability of the non-trivial uniform steady-states}
Next, we examine the stability of the uniform non-trivial states which is governed by the determinant of the Jacobian ${\bf J}^*|_{q=0}$ in (\ref{J})
\begin{equation}
\det= \nu c_0^*[d+ { \nu}c_0^*+{2} ({\mu} +{ \nu} A^*){\cal P}'(A^*)].
\end{equation}
With ${\cal P}(A)$ given by (\ref{PA}), $c_0^*=2 {\cal P}(A^*)-1$ and $X\equiv \gamma A^*$, one has
\begin{equation}
\det=d^2 {\tilde \nu}c_0^*[1+ {\tilde \nu}c_0^*-2mp ({\tilde \mu}+ {\tilde \nu} X)\frac{X^{m-1}}{(1+X^m)^2}]. \label{detX}
\end{equation}
$X$ satisfies  $F(X)=0$ from which one can express $X$ in terms of $Y\equiv X^m$ as
\begin{equation}
X=\frac{2{\tilde \mu}(1-p+Y)}{(1-{\tilde \nu})Y+1-{\tilde \nu}+2p{\tilde \nu}}.\label{XY}
\end{equation}
Upon substituting (\ref{XY}) in (\ref{detX}) and after some algebra, the determinant can be written
as
\begin{equation}
\det=\frac{d^2 {\tilde \nu}c_0^*(Y+1)}{1-p+Y}\left\lbrace (1-{\tilde \nu})Y^2-[mp(1+{\tilde \nu})+(2-3p){\tilde \nu}+p-2]Y+(1-p)(1-{\tilde \nu} +2p {\tilde \nu})\right\rbrace.
\end{equation}
It is clear that det contains the same quadratic factor in $Y$ as in (\ref{Y2}) and hence the boundaries for the emergence of new pair of roots, ${\tilde \mu}_t^{\pm}$ in (\ref{mutpm}) are also the stability boundaries $\det=0$.

In addition,  for the special value of  ${\tilde \mu}={\tilde \mu}_0(p)\equiv (2p-1)^{1\over m}$ (which is the stability boundary of the trivial USS (\ref{mu0p})), $F(X)$ in (\ref{FX}) becomes
\begin{equation}
F(X)= (1-{\tilde \nu}) X^{m+1}-{2{\tilde \mu}}X^m+(1+{\tilde \nu}{\tilde \mu}^m)X -{\tilde \mu}(1-{\tilde \mu}^m).\label{FX2}
\end{equation}
It is easy to verify directly in (\ref{FX2}) that $F(X={\tilde \mu})=0$, 
and hence $X={\tilde \mu}$ is always a  non-trivial root  on the ${\tilde \mu}={\tilde \mu}_0(p)$ curve for arbitrary values of ${\tilde \nu}$. Furthermore, since $c_0^*=2p/(1+X^m)-1=0$, thus the corresponding det vanishes (see Eq. (\ref{delC0})) on the ${\tilde \mu}_0(p)$ line, regardless of the values of ${\tilde \nu}$. This result also agrees with (\ref{mu0}). Hence the ${\tilde \mu}_0(p)$ line is also the stability boundary for one of the non-trivial roots.


\section*{Appendix II: Numerical methods}
Since the dynamics of the regulatory molecules  is much faster than that of  the tissue growth rate, it is  reasonable to assume the quasi-static condition (with $V=0$ and $\partial_tA=0$, but $\partial_z V\neq 0$) for the dynamics of $A$ in the numerical computation. By choosing the time and space in  units of $1/d$ and $\sqrt{D/d}$ respectively, from (\ref{PDE2}) the steady-state distribution of $A$ in a fixed spatial domain of $[0, L_{m}]$ obeys
\begin{equation}
[\partial^2_z -(1+{\tilde \nu} c_0(z))]A +s(z)=0\label{PDEA}
\end{equation}
 where $s(z)=\tfrac{\mu}{d}(1-c_0(z))$, 
  together with the no flux boundary conditions at $z =0$ and $z= L_{m}$, 
 \begin{equation}
\partial_z A|_0=\partial_z A|_{L_{m}}=0.\label{BCA}
 \end{equation}
 (\ref{PDEA}) with boundary conditions (\ref{BCA})
  can be solved  using the Green's function approach. In particular, for the case of $c_0\to 0$ or ${\tilde \nu}=0$, the Green's function can be  derived analytical to be
\begin{equation}
G(z,z')=  \frac{\cosh z_<\cosh(L_{m}-z_>)}{\sinh L_{m}}\label{Gzz}
\end{equation} 
where $z_>$ ($z_<$) denotes the greater (lesser) of $z$ and $z'$. The quasi-static solution of $A$ is then given by
\begin{eqnarray}
A_{\nu\to 0}(z)= \frac{\mu/d}{\sinh L_m} &( & \int_0^z dz'   \cosh z'\cosh(L_m-z)[1-c_0(z')]\nonumber \label{Az}\\
& &+ \int_z^{L_m} dz'   \cosh z\cosh(L_m-z')[1-c_0(z')]).
\end{eqnarray}
For ${\tilde \nu} >0$ and $c_0>0$, (\ref{PDEA}) with boundary conditions (\ref{BCA})
 are solved numerically for given $c_0(z)$.
Space is discretized into grid points f mesh size $\Delta x$ in the range of 0.1 to 0.2. Since (\ref{PDEA}) is a linear ordinary differential equation and hence  using the finite different method, the discretized system can be solved conveniently by linear algebra  with a home-made code using the LAPACK package\cite{Lapack}.  For  given initial profile of $c_0(z)$ and $A(z)$, which are usually taken to be step-functions, time is marched forward with a fixed time-step $\delta t= 10^{-3} ~ 10^{-4}$. At a given time $t$ with given $c_0(z,t)$ and $L_m(t)$, $A(z)$ is numerically solved from (\ref{PDEA})   on the grid points. Then $L_m(t+\delta t)$ at the next time step is advanced forward using (\ref{Vz2}) and (\ref{LmODE}), and $c_0(z,t+\delta t)$ computed from (\ref{PDE3}).
  From the initial $c_0(z,0)$, the above calculations are repeated for each forward time step and hence the numerical solutions for the dynamics can be obtained.


\begin{thebibliography}{99}
\bibitem{SCTD}Suppose SC proliferates or differentiates to SC$\to$SC+SC, SC$\to$SC+TD and SC$\to$TD+TD with probabilities $r$, $1-r-s$ and $s$ respectively, then it is easy to show that now the rhs of the first  and second equations in (\ref{PDEG1}) become $\nu c_0(r-s)$ and $\nu c_0(1-r+s)$ respectively. Hence by re-defining $2{\cal P}-1\equiv r-s$, one effectively has the same model depicted as in Fig. \ref{schematic}a. Hence even if the asymmetric division SC$\to$SC+TD is included, the system can still be described by the form  in (\ref{PDEG1}).
\end{thebibliography}
\end{document}